\documentclass[fleqn,usenatbib,useAMS]{mnras}

\usepackage{newtxtext,newtxmath}

\usepackage[T1]{fontenc}
\usepackage{ae,aecompl}

\usepackage{graphicx}	
\usepackage{amsmath}	
\usepackage{amssymb}	
\usepackage{xspace}
\usepackage{gensymb}
\hypersetup{draft}

\defcitealias{Poggianti2017a}{Paper I}
\defcitealias{Bellhouse2017}{Paper II}
\defcitealias{Fritz2017}{Paper III}
\defcitealias{Gullieuszik2017}{Paper IV}
\defcitealias{Moretti2018}{Paper V}
\defcitealias{Vulcani2017c}{Paper VIII}
\defcitealias{Vulcani2018}{Paper VII}
\defcitealias{Jaffe2018}{Paper IX}
\defcitealias{Vulcani2018c}{Paper XIV}
\defcitealias{Vulcani2018b}{Paper XII}

\newcommand{\Ha}{H$\alpha$\xspace}
\newcommand{\Hb}{H$\beta$\xspace}

\newcommand{\kms}{$\rm km \, s^{-1}$\xspace}
\newcommand{\kube}{{\sc kubeviz}\xspace}
\newcommand{\sinopsis}{{\sc sinopsis}\xspace}

\title[Physical processes in filaments]{GASP. XVI. Does cosmic web enhancement turn on star formation in galaxies?
}

\author[B. Vulcani et al.]
{Benedetta Vulcani,$^{1}$\thanks{E-mail: benedetta.vulcani@inaf.it (BV)}
Bianca M. Poggianti,$^{1}$
Alessia Moretti,$^{1}$
Marco Gullieuszik,$^{1}$
\newauthor
Jacopo Fritz,$^{2}$
Andrea Franchetto,$^{3,1}$
Giovanni Fasano,$^{1}$
Daniela Bettoni,$^{1}$
Yara L. Jaff\'e,$^{4}$\\
$^{1}$INAF- Osservatorio astronomico di Padova, Vicolo Osservatorio 5, IT-35122 Padova, Italy\\
$^{2}$Instituto de Radioastronom\'ia y Astrof\'isica, UNAM, Campus Morelia, A.P. 3-72, C.P. 58089, Mexico\\
$^{3}$Dipartimento di Fisica \& Astronomia ``Galileo Galilei'', Universit\`a di Padova, vicolo dell' Osservatorio 3, IT 35122, Padova, Italy\\
$^{4}$Instituto de F\'isica y Astronom\'ia, Universidad de Valpara\'iso, Avda. Gran Breta\~na 1111 Valpara\'iso, Chile}

\date{Accepted XXX. Received YYY; in original form ZZZ}

\pubyear{2018}

\begin{document}
\label{firstpage}
\pagerange{\pageref{firstpage}--\pageref{lastpage}}
\maketitle

\begin{abstract}
Galaxy filaments are a peculiar environment, and their impact on the galaxy properties is still controversial. Exploiting the data from the GAs Stripping Phenomena in galaxies with MUSE (GASP), we provide the first characterisation of the spatially resolved properties of galaxies embedded in filaments in the local Universe. The { four} galaxies we  focus on show peculiar ionised gas distributions:  \Ha clouds have been observed  { beyond four times the effective radius}. The gas kinematics, metallicity map and the ratios of emission line fluxes confirm that they do belong to the galaxy gas disk, the analysis of their spectra shows that very weak stellar continuum is associated to them. Similarly, the star formation history and luminosity weighted age maps point to a recent formation of such clouds. The clouds are powered by star formation, and are characterised by intermediate values of dust absorption. We hypothesise  a scenario in which the observed features are due to ``Cosmic Web Enhancement'': we are most likely witnessing galaxies passing through or flowing within filaments that assist the gas cooling  and increase the extent of the star formation in the densest regions in the circumgalactic gas.  Targeted simulations are mandatory to better understand this phenomenon. 
\end{abstract}




\section{Introduction}
\label{sec:intro}

The properties of galaxies are directly affected by their host environment. In the local universe, red, passive, early-type galaxies are preferentially found in dense regions and galaxy clusters while blue, star-forming, late-type galaxies dominate in less dense, field environments \citep[e.g.,][]{Dressler1980, Kauffmann2004, Balogh2004}.

Studies of the effect of large-scale structure on galaxy properties are usually mostly confined to field versus clusters. However, the intermediate environments such as galaxy groups, cluster outskirts and filaments are equally important \citep[e.g.,][]{Kodama2001}, as they host the vast majority of galaxies in the local universe \citep[e.g.,][]{Jasche2010, Tempel2014_f, Tempel2014_g, Cautun2014}. 

In particular, filaments that connect groups and clusters of galaxies may contain up to 40 per cent of the matter in the Universe \citep{Forero2009, Jasche2010}. 
Theoretical studies \citep[e.g.,][]{Cen1999} have suggested that about half of the warm gas in the Universe, presumably accounting for the low-redshift missing baryons \citep{Fukugita1998, Viel2005}, is hidden in filaments. Recently, \cite{Nicastro2018} 
 observed highly ionized oxygen systems in regions characterized by large galaxy over-densities, supporting the prediction of warm gas in the extragalactic universe.

Trying to dissect the role of these environments on galaxy properties is therefore of extreme importance to shed light on the processes that regulate galaxy evolution. 

Several mechanisms have been proposed to govern galaxy properties. Dark matter filaments can trap and compress gas, shock heating the accreted gas at the boundary of filaments. This gas then cools rapidly and condenses into filaments centre. Filaments can therefore assist gas cooling and enhance star formation in their haloes \citep{Liao2018}. 
In filaments, mild galaxy-galaxy harassment and interactions \citep{Lavery1988, Moore1996, Coppin2012} are favored. Their environment  is colder than clusters: the typical temperature of filaments is $\sim 10^5-10^7$ K \citep[e.g.,][]{Cen2006, Werner2008, Zappacosta2002, Nicastro2005}{, even though filaments must also contain cool gas (T$\sim 10^4$ K), as predicted by Lyman alpha forest observations \citep[e.g.,][]{Kooistra2017}}. Therefore, galaxies in filaments can still hold their gas content to form stars. Nonetheless, for low mass galaxies ($M_\ast<10^{10}M_\odot$), simulations show that filaments falling onto clusters are able to produce increased stripping of hot gas even beyond a distance of 5$r_{200}$  from a galaxy cluster centre  and that this is predominant at low-redshift \citep{Bahe2013}, suppressing the fuel for star formation. 
In filaments, ram-pressure stripping \citep{Gunn1972} is known as cosmic web stripping  and is due to the interaction of the galaxies and the filaments, and might also play a role, especially for low mass galaxies, whose shallow potential wells can provide a relatively small restoring force from the ram-pressure force of the IGM in filaments \citep[e.g.,][]{Benitez2013}. 
However, this effect has been never observed and the fate of the gas that remains in the galaxy or is accreted later is not clear. The cosmic web stripping is so far a purely hydro-dynamical effect that requires simulations of large volumes able to resolve properly both the cosmic web and the internal halo properties.

Another process that   is also expected to be quite effective in filaments is  gas accretion, which increases the availability of cold gas for galaxies  inducing an enhancement of the star formation \citep[e.g.,][]{Darvish2014}.

Several works have shown that  filaments  affect the evolution of the integrated properties of galaxies \citep[e.g.,][]{
Koyama2011, Geach2011, Sobral2011, Mahajan2012, Tempel_Libeskind2013, Tempel2013, Zhang2013, Pintos2013, Koyama2014, Santos2014, 
Malavasi2017, 
Mahajan2018} and the distribution of satellites around galaxies \citep{Guo2014}, at any redshift, but results are still controversial. 
Overall, filament galaxies tend to be more massive, redder, more gas poor  and have earlier morphologies  than galaxies in voids \citep{Rojas2004, Hoyle2005, Kreckel2011, Beygu2017, Kuutma2017}.  On the other hand, some studies  have reported an increased fraction of star-forming galaxies \citep{Fadda2008, Tran2009, Biviano2011, Darvish2014}, and higher metallicities and lower electron densities \citep{Darvish2015} in filaments with respect to field environments.

{ Differences in the results might also be due to the different techniques adopted by different teams to define filaments. Indeed, due to the observational biases in large galaxy surveys and unvirialized nature of the large-scale structures, their characterization  is a nontrivial task and many assumptions come into play \citep[e.g., ][]{Biviano2011, Tempel2014_f, Poudel2017}.}

{ From the theoretical point of view,} 
\cite{Gay2010} have investigated the influence of  filaments on the spectroscopic properties of galaxies, using the MareNostrum simulation. They found that the large-scale filaments are only dynamical features of the density field, reflecting the flow of galaxies accreting on clusters; the conditions in the filaments are not dramatic enough to influence strongly the properties of the galaxies it encompasses. 
On the other hand, \cite{Aragon2016} showed that the star formation quenching in galaxies can be explained as the influence of filamentary environment. 
So far,  no studies have investigated how spatially resolved properties could be affected by filaments,  from neither an observational nor theoretical point of view, except for our attempt in \citet[Paper XII]{Vulcani2018b}.

In this paper we present the analysis of { four} field spiral galaxies in the local universe showing asymmetric features { and an extended \Ha distribution, proxy for extended  H{\sc ii} regions,} that we will argue are most likely due to the effect of the hosting filaments. 

{ 
 H{\sc ii} regions signifying the presence of  ionising OB stars are usually found in the luminous inner regions of galaxies \citep[see, e.g.,][]{Martin2001}.
The  evidence of star formation in outer disks, instead,  raises new questions about the nature of star formation in diffuse environments.
Indeed, outer disks are usually considered inhospitable environments for star formation. In fact, a deviation in the  Kennicutt-Schmidt Law \citep{Kennicutt1998b, Kennicutt1989} has been observed at a gas surface density of 3-5 M$_\odot \,  pc^{-2}$,
where the 
\Ha intensity suddenly drops  \citep[but see][who suggest this is merely a stochastic effect]{Boissier2007}. 
This is generally interpreted as a threshold density for star formation \citep{Kennicutt1989, Martin2001}, most likely due to a transition between dynamically unstable and stable regions of the galaxy \citep[e.g.,][]{Toomre1964} or to a phase transition of the gas \citep[e.g.,][]{Elmegreen1994, Schaye2004, Krumholz2009}.

However, \Ha knots at large radii have been observed in a few galaxies \citep{Kennicutt1989, Martin2001, Ferguson1998} and $\sim$30\% of disk galaxies have UV emitting sources beyond their optical disks \citep{Thilker2005, Thilker2007, GildePaz2005, Zaritsky2007, Christlein2008}. These complexes are often coincident with local H{\sc i} over-densities \citep{Ferguson1998}. In M83 and NGC 4625, the UV knots have been identified as low-mass stellar complexes and, if visible in the \Ha, are generally ionised by a single star \citep{GildePaz2007}. These knots are dynamically cold and rotating, indicating that outer disk complexes are extensions of the inner disk \citep{Christlein2008}.

Isolated H{\sc ii} regions have also  been discovered in the extreme outskirts of galaxy halos in the Virgo Cluster \citep{Gerhard2002, Cortese2004}, in gaseous tidal debris \citep{RyanWeber2004, Oosterloo2004} and in between galaxies in galaxy groups \citep{Sakai2002, Mendes2004}. These appear as tiny emission-line objects in narrow-band images at projected distances up to 30 kpc from the apparent host galaxy.
These regions sometimes are associated with previous or ongoing galaxy interactions \citep{Thilker2007, Werk2008}.

\Ha radiation has also been observed to be emitted by the gaseous halos of nearby galaxies \citep{Zhang2018}. This emission is extremely faint (flux${\rm \ll 10^{-17} erg/cm^2/s/}$\AA{}) and has been observed up to several hundreds of kpc from the main galaxy. 

An explanation for the existence of these outer knots could be that at some sites the gas density may exceed a star formation threshold locally, allowing stars to form beyond the radius where the azimuthally averaged gas density is at or below a threshold density \citep{Kennicutt1989, Martin2001, Schaye2004, Elmegreen2006,  GildePaz2007}. 

All of above studies are based on traditional observational techniques, such as narrow-band imaging and Fabry-Perot staring technique. These techniques only permit the detection and basic characterization of the H{\sc ii} regions, without giving spatially resolved information on the chemical composition and age of the regions.

The galaxies we discuss in this paper instead are drawn from a Integral Field Spectrographs (IFS) survey that has been  designed to focus on the galaxy external regions and allows us to perform a detailed analysis of the galaxy outskirts. 
}

GASP\footnote{\url{http://web.oapd.inaf.it/gasp/index.html}} (GAs Stripping Phenomena in galaxies with MUSE), an  ESO Large programme that exploits the integral-field spectrograph MUSE mounted at the VLT with the aim to characterise where, how and why gas can get removed from galaxies in different environments. A complete description of the survey strategy, data reduction and analysis procedures is presented in \cite[][Paper I]{Poggianti2017a}. 
First results on single  objects in clusters are discussed in \citealt[(Paper II)]{Bellhouse2017}; \citealt[(Paper III)]{Fritz2017}; \citealt[(Paper IV)]{Gullieuszik2017}; \citealt[(Paper V)]{Moretti2018}; and in lower-density environments in \citealt[(Paper VIII)]{Vulcani2017c}; \citealt[(Paper VII)]{Vulcani2018}; \citetalias{Vulcani2018b}.

 GASP includes a sample of galaxies  selected for presenting a B-band morphological asymmetry suggestive of unilateral debris \citep{Poggianti2016} plus a subset of undisturbed galaxies, used as control sample. 

Throughout all the papers of the GASP series, we adopt a \cite{Chabrier2003} initial mass function (IMF) in the mass range 0.1-100 M$_{\odot}$. The cosmological constants assumed are $\Omega_m=0.3$, $\Omega_{\Lambda}=0.7$ and H$_0=70$ km s$^{-1}$ Mpc$^{-1}$. 

\section{Data}
\subsection{The target selection}
{ In this paper, unless otherwise stated, we consider only the GASP galaxies selected from the field sample. All galaxies are drawn from the Millennium Galaxy Catalog \citep{Liske2003, Driver2005} and selected from the PM2GC \citep{Calvi2011}.}
{
We exclude from the GASP sample interacting  \citepalias[e.g.][]{Vulcani2017c} and passive \citepalias[e.g,][]{Vulcani2018b} galaxies, counter-rotating disks \citepalias[e.g.][]{Vulcani2018b} and galaxies with a central \Ha hole (Moretti et al. in prep.).

We compute the maximum extension of the \Ha distribution in units of effective radius ($r_e$). Specifically, we measure the radius containing 99\% of the \Ha flux having a S/N>3. Details on the \Ha images used for selecting the galaxies are given in Sec 2.2 and 2.3. 

The effective radius, along with the inclination and the position angle of the galaxies, were obtained from the analysis of the I-band images achieved from the integrated MUSE datacubes on the Cousins I-band filter response curve, as explained in Franchetto et al. (in prep.). Briefly, they were obtained using {\tt ellipse} \citep{Jedrzejewski1987} 
of the software IRAF that allows an isophotal segmentation of the galaxy and 
 draws the luminosity growth curve
\begin{equation}
    L(R)=2\pi\,\int_{0}^{R}I(a)\,(1-\varepsilon(a))\,a\,da,
\end{equation}
where $I(a)$ is the surface brightness profile, $\varepsilon(a)$ is the isophotal ellipticity profile and $a$ is the semi-major-axis of the elliptical isophotes.

Taking advantage of the ample sky coverage of the GASP data, we extended the fitting up to the most external part of the galaxies to probe the behaviour of the surface brightness at large radii.
 Sources extraneous to the galaxy, brighter knots -often located along the spiral arms- and bad pixels were masked out to prevent erroneous measurements.
Although the I-band image is obtained from sky-subtracted MUSE datacube,  it presents residual sky intensity comparable to the intensities of last fitted isophotes. Thus, we subtracted the value of the intensity of the last isophote to the image and proceeded with the computation of the luminosity growth curve. 

Assuming that the galaxy regions over the largest isophote negligibly contribute to the total galaxy luminosity, we approximated $L_{\rm tot}\approx L(a_{\rm max})$ - with $a_{max}$ semi-major-axis of the  largest  isophote - and we estimated the effective radius as the radius $R_{\rm e}$ such as $L(R_{\rm e})/L_{\rm tot}=0.5$. 

From the surface brightness profile we selected the isophotes that trace the stellar disk to measure their mean  position angle ($PA$), the mean ellipticity ($\varepsilon$) and corresponding errors.

\begin{figure}
\centering
\includegraphics[scale=0.4]{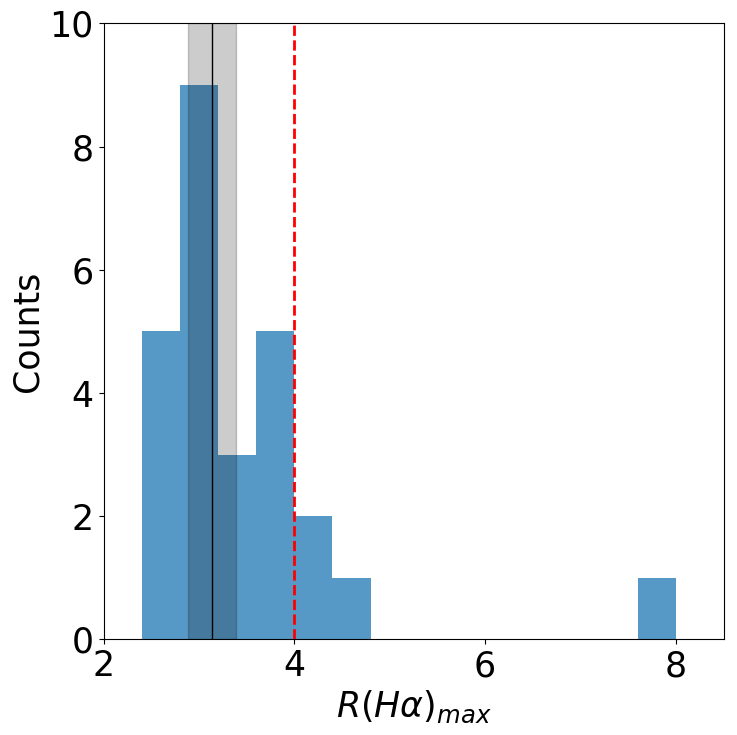}
\caption{{Maximum extent of \Ha in units of $r_e$ (R(\Ha)$_{max}$) distribution in the GASP field sample. Black line and shaded area show the median value and its error. The red line shows the threshold used to select galaxies in this work. } \label{fig:rmax} }
\end{figure}

\begin{figure}
\centering
\includegraphics[scale=0.58]{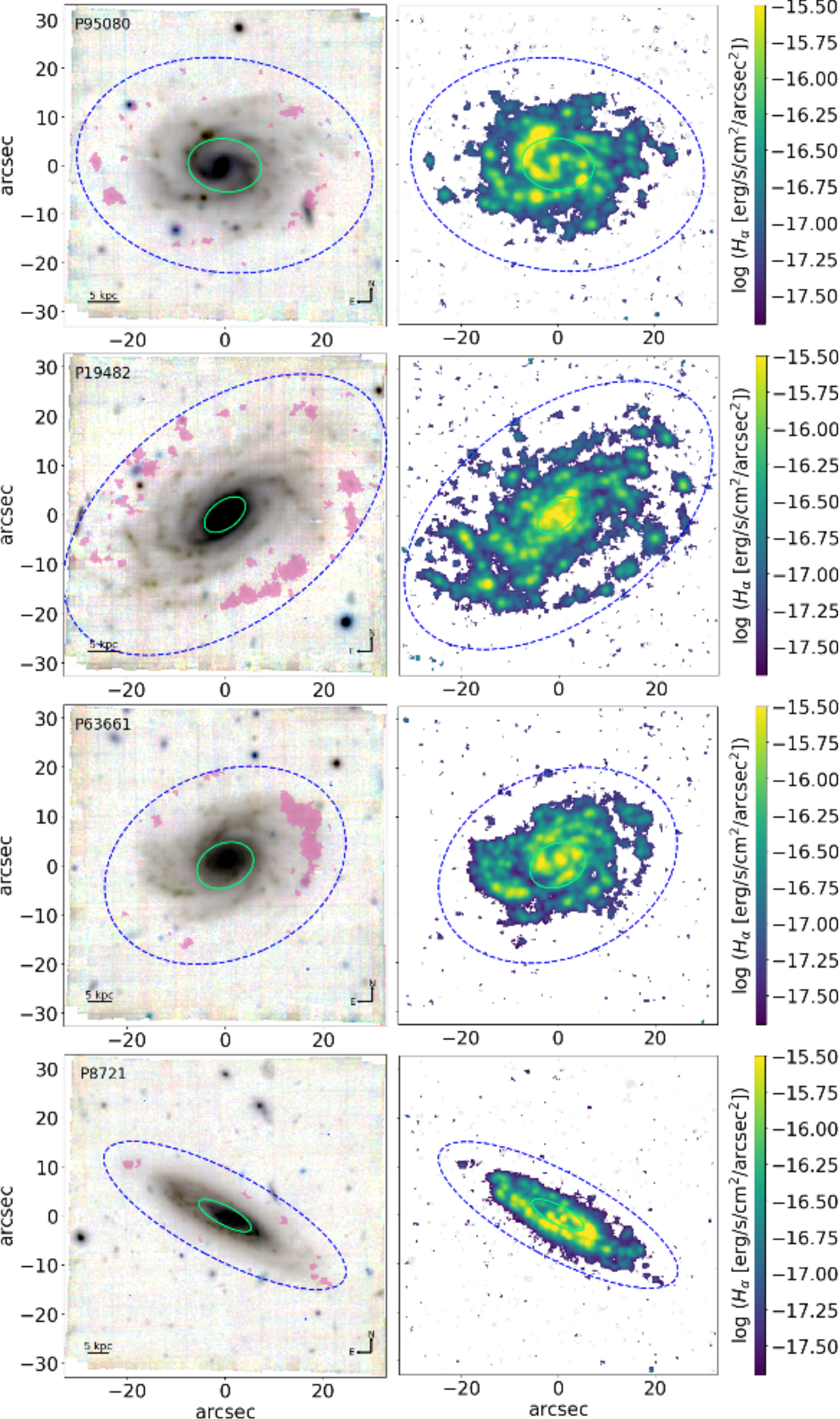}
\caption{{ RGB (left) and \Ha (right) images of the targets. From  top to bottom P95080, P19482, P63661, and  P8721 are shown. The reconstructed $g$, $r$, $i$  filters from the MUSE cube have been used. North is up, and east is left.  Color map is inverted for display purposes.  In all the plots, the green ellipses show the  $r_e$, the dashed blue ellipses show the maximum radius at which \Ha is detected (see text for details). {  Purple areas show the detached clouds (see text for details).}   Asymmetries in the star and gas distributions are seen in all galaxies, with one side of the galaxies extending more than the other.  All these galaxies have \Ha extending beyond 4$r_e$ and show patchy \Ha distribution. } \label{fig:rgb_image} }
\end{figure}

The \Ha disk extension of the sample is shown in Fig. \ref{fig:rmax}. 

The median \Ha disk extension is 3.1$\pm$0.2
times $r_e$. We then selected the galaxies with maximum \Ha extension larger than four $r_e$, corresponding to 90th percentile.  Four galaxies passed the selection and they are listed in Table \ref{tab:gals}, which  } summarises some important information that will be further used and discussed throughout the paper. 

\begin{figure}
\centering
\includegraphics[scale=0.57]{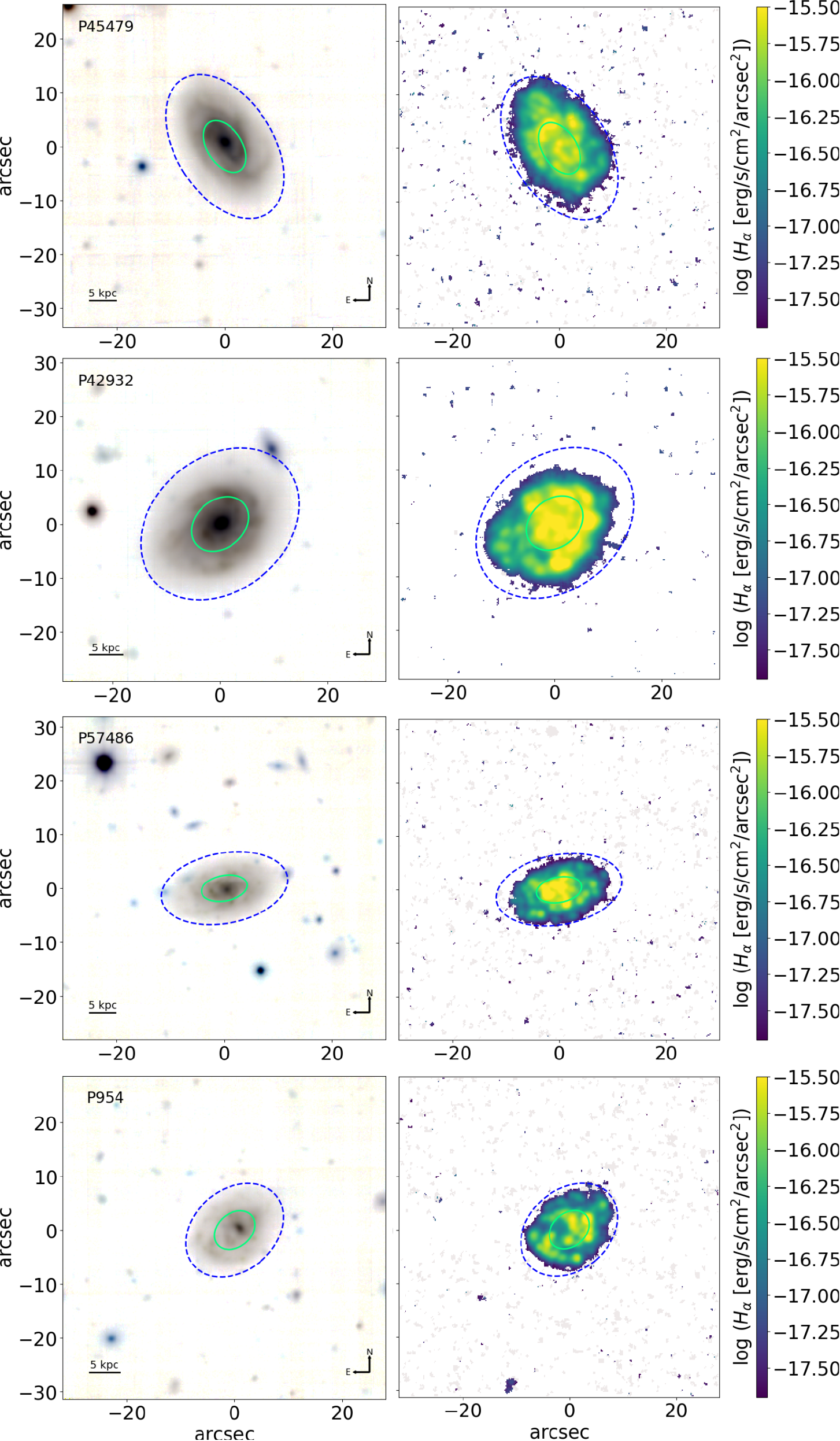}
\caption{{ Same as Fig.\ref{fig:rgb_image}, but for four galaxies representative of the GASP control sample. In these galaxies \Ha extends at most to 3$r_e$ and the galaxy boundaries are not jagged. } \label{fig:rgb_image_control} }
\end{figure}

Figure \ref{fig:rgb_image} shows the color composite images of the  targets, obtained combining the reconstructed $g-$, $r-$ and $i-$filters from the MUSE datacube 
 along with the \Ha maps. Overplotted in green are the $r_e$, while overplotted in blue are the maximum radii at which \Ha is detected. For comparison, Figure \ref{fig:rgb_image_control} shows the color composite images and \Ha maps of four representative galaxies of the control sample. Figures 1 and 6 in Vulcani et al. (submitted, Paper XX), show the images for all the galaxies in the GASP control sample that will be also used in this paper (sec 3.5). The presence of detached clouds { (highlighted in purple in Fig. \ref{fig:rgb_image})} in the \Ha disk of the selected galaxies is astonishing, especially if compared with the absence of the same features among the control sample galaxies.
 The clouds extend beyond the spiral arms of the galaxies, suggesting they might not strictly related to them. {  A quantitative identification of the clouds is deferred to the next Section.}

\begin{table*}
\caption{Properties of the targets. For each galaxy, the ID, redshift, coordinates, total stellar mass, effective radius,  maximum extension of \Ha,  position angle, ellipticity and physical scale are given. \label{tab:gals}}
\centering
\begin{tabular}{lrrrrrrrrr}
\hline
  \multicolumn{1}{c}{ID} &
  \multicolumn{1}{c}{z} &
  \multicolumn{1}{c}{RA} &
  \multicolumn{1}{c}{DEC} &
  \multicolumn{1}{c}{$\log M$}  &
  \multicolumn{1}{c}{$r_e$} &
  \multicolumn{1}{c}{$R(H\alpha)_{max}$} &
  \multicolumn{1}{c}{PA} &
  \multicolumn{1}{c}{$\epsilon$} &
   \multicolumn{1}{c}{phys. scale}
  \\
  \multicolumn{1}{c}{} &
  \multicolumn{1}{c}{} &
  \multicolumn{1}{c}{(J2000)} &
  \multicolumn{1}{c}{(J2000)} &
  \multicolumn{1}{c}{($M_\ast/M_\odot$)}&
  \multicolumn{1}{c}{($^{\prime\prime}$)} &
  \multicolumn{1}{c}{($r_e$)} &
  \multicolumn{1}{c}{(deg)} &
  \multicolumn{1}{c}{} &
   \multicolumn{1}{c} {kpc/$^{\prime\prime}$}  \\

\hline
P90580 & 0.04038 &198.03625 &-0.23903 & 9.98 &7.5$\pm$0.8 & 4.1&83 & 0.32 & 0.7985\\
P19482 & 0.04063 & 170.63021 & -0.01711 & 10.27 &4.8$\pm$0.4 &7.9&127 &0.42& 0.8030\\
P63661 & 0.05516 & 218.09081 &0.17823 & 10.26 & 6.0$\pm$0.6 &4.3&-63 & 0.41 &1.0718  \\
P8721 &0.06477 & 158.53624 & 0.00101 & 10.75 & 6.0$\pm$0.2 &4.6&62&  0.67  &1.2443 \\
\hline\end{tabular}
\end{table*}

\subsection{Observations and data reduction}
All the GASP targets were observed in
service mode with the MUSE spectrograph, mounted at the Nasmyth focus of the UT4 VLT, at Cerro Paranal in Chile. Each galaxy was observed with clear conditions; the seeing remained below 0$\farcs$9 during observations. For each galaxy, a total of four 675 seconds exposures were taken with the Wide Field Mode. AS far as the galaxies discussed in this paper is concerned, P95080 was observed on 2017, February 4; { P15982 on 2017, May 5;} P63661 on 2017, May 1 and  P8721 on 2016, January 9.

The data reduction process for all galaxies in the GASP survey is presented in  \citetalias{Poggianti2017a}.  
For all galaxies, we average filtered the datacubes in the spatial direction with a 5$\times$5 pixel kernel, corresponding to 1$^{\prime\prime}$ \citepalias[see][for details]{Poggianti2017a}. At the redshifts of the galaxies presented here,  1$^{\prime\prime}$ that corresponds to 0.8-1.2 kpc, depending on the redshift of the target.

\subsection{Methods}\label{sec:analysis}
 \citetalias{Poggianti2017a} extensively presents the methods  used to analyse galaxies within the GASP program. Here we just recall the basic procedures and references useful for the following analysis. 
In brief, we corrected the MUSE reduced datacubes for extinction due to our Galaxy and then we measured (1) the total fluxes and kinematic properties of the gas, by running the  \kube  \citep{Fossati2016} code; (2) the kinematic properties of the stars, by running the  Penalized Pixel-Fitting (pPXF) software \citep{Cappellari2004}, which works
in Voronoi binned regions of given S/N \citep{Cappellari2003} and smoothed using the two-dimensional local regression techniques (LOESS) as implemented in the Python code developed by M. Cappellari;\footnote{\url{http://www-astro.physics.ox.ac.uk/~mxc/software}} (3) the properties of the stellar populations, such as star formation histories, luminosity and mass weighted ages, surface mass densities, by running the spectral fitting code  \sinopsis \citepalias{Fritz2017};  (4) the  dust extinction  A$_V$  from the 
absorption-corrected
Balmer decrement assuming an intrinsic \Ha/H$\beta$ ratio equal to 2.86 and adopting the \cite{Cardelli1989} extinction law; and
(5) ionised gas metallicity, by running a modified version of the pyqz Python
\citep{Dopita2013} v0.8.2 (F. Vogt 2017, private communication).

Further details will be discussed in the next section, where needed.

\section{RESULTS}\label{sec}
In this section we characterise separately each of the targets. In the following sections we will highlight what these galaxies have in common and look for the reasons of such similarities. 
\subsection{P95080}

\begin{figure*}
\centering
\includegraphics[scale=0.5]{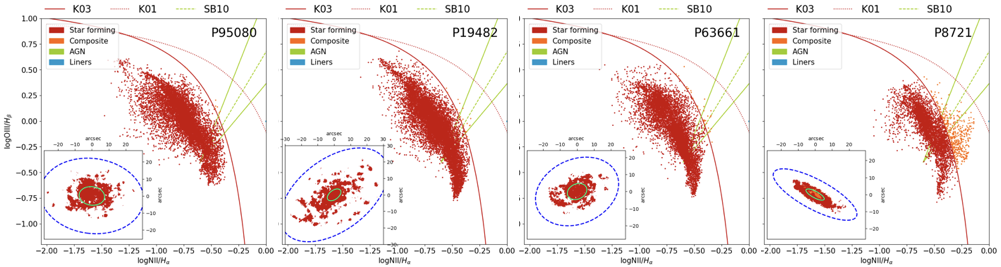}
\caption{BPT line-ratio  diagram for [OIII]5007/H$\beta$ vs [NII]6583/\Ha for the three galaxies. Lines  are from \citet[][K03]{Kauffmann2003}, \citet[][K01]{Kewley2001} and \citet[][SB10]{Sharp2010} to separate Star-forming, Composite, AGN and LINERS. In the inset the BPT line-ratio map is shown. Only spaxels with a $S/N> 3$ in all the emission lines involved are shown. \label{fig:BPT} }
\end{figure*}

\begin{figure*}
\centering
\includegraphics[scale=0.51,clip, trim=0 0 0 0]{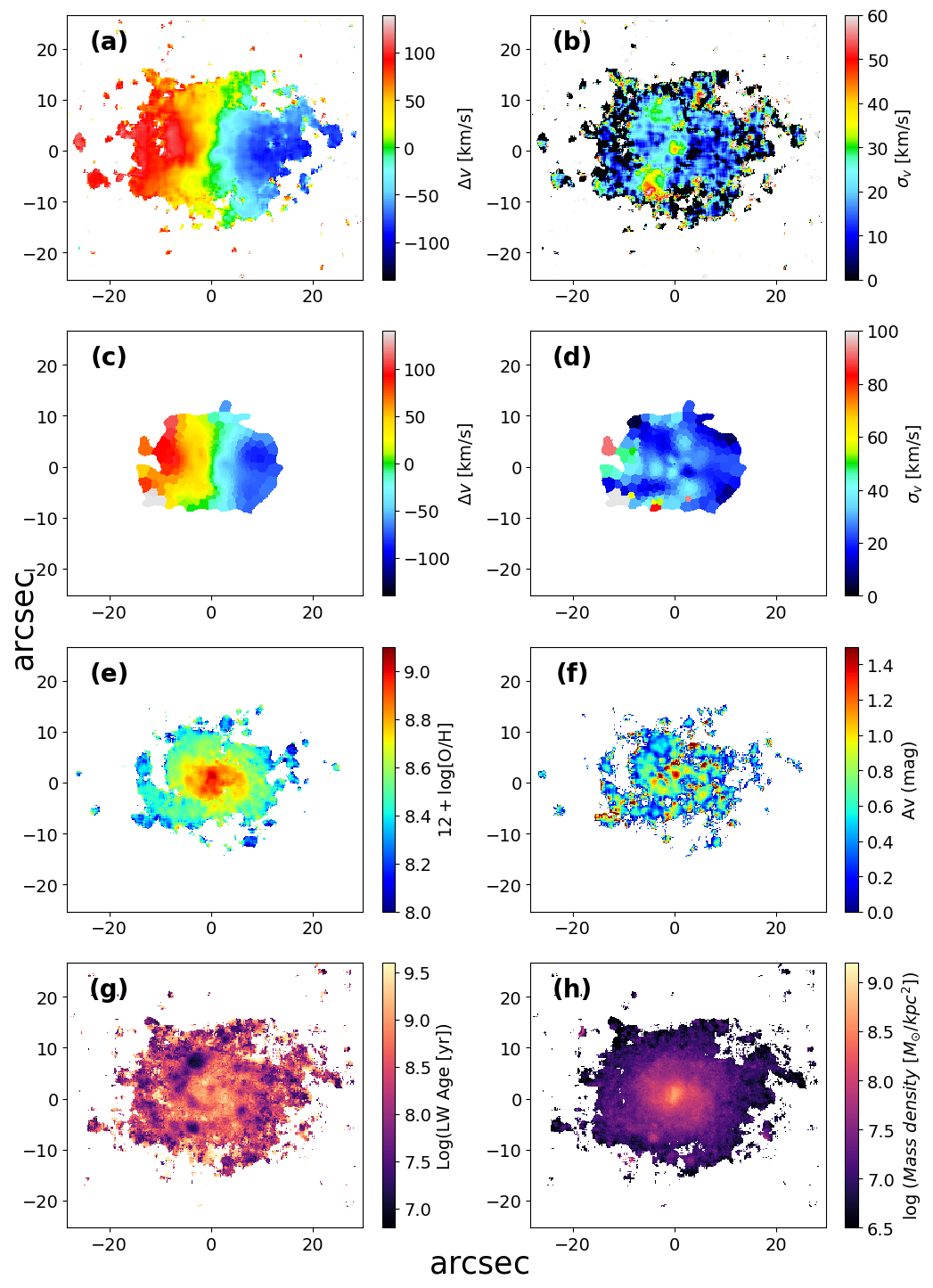}
\caption{P95080. The different panels show the  MUSE map of \Ha (a), the \Ha velocity  (b) and  velocity dispersion (c) maps, stellar velocity (d) and stellar  velocity dispersion (e) maps,  metallicity map for the ionised gas (f), $A_V$ maps (g), luminosity weighted age (h) and stellar mass density (i) maps. More details are given in the text. 
In all plots, (0, 0) is the center of the MUSE image. 
\label{fig:P95080} }
\end{figure*}

P95080, seen in the { top panels} of  Fig.\ref{fig:rgb_image},  is a spiral galaxy with a moderate inclination and a possibly a bar. 
The right panel shows the MUSE map for \Ha, uncorrected for intrinsic dust extinction, but corrected for stellar absorption and Galactic extinction.  We plot only the spaxels with \Ha S/N$> 3$. The \Ha distribution is quite patchy in the central regions of the galaxy, where  many peaks  are visible. It emerges that the ionised gas is characterised  a number of clouds detached from the main body that extend beyond the visible light. These surround the galaxy without having a preferred orientation. As these clouds might be only due to spurious spaxels, we decided to plot only the spaxels that, in addition to having \Ha S/N$> 3$, { are surrounded by spaxels with measured velocity at S/N>3. Specifically, we build a 3$\times$3 matrix centered on each spaxel and we keep only those spaxels that are surrounded by at least 7 (out of 9) spaxels with non-zero velocity}. In addition, for the spaxels in the outskirts of the galaxy and in possible isolated clouds, we will plot only the spaxels whose velocity is within $3\sigma$ of the mean velocity of the  the galaxy{ , considering separately the approaching and receding sides}.
This approach helps to remove possible spurious signal.

Given the redshift of P95080, the sky line at $\lambda$=6830\AA{} falls very close to \Ha. Therefore, only for this galaxy, we also  exclude all the spaxels in the clouds whose velocity is within $\pm 50$ \kms the velocity of the sky line. 
We will clean also the following plots adopting the same approach.  
We are therefore confident that the clouds we detect are real and due to some specific physical process. {  To identify the clouds using a quantitative metrics, we select all \Ha regions in the luminosity range $\rm 10^{-17.7}-10^{-15.5} \, erg/s/cm^2/acrsec^2$ that have no pixels in common with the main body of the galaxy,  have a size larger than 10 pixels and are within R(\Ha)$_{max}$. With this approach, we identify 32 clouds.} In P95080, the measured extension of the  \Ha disk is 4.1$\times$ the extension of the stellar disk, defined by $r_e$.

We note that the \Ha images shown in Fig. \ref{fig:rgb_image_control} have been produced following the same procedure, therefore clouds of similar size and intensity would be detected. {  In contrast, the aforementioned approach does not identify any clouds in the galaxies belonging to the control sample. }

The  map of \Ha, when in combination  to those of  \Hb, [OIII] 5007 \AA{}, [OI] 6300 \AA{}, \Ha, [NII] 6583 \AA{}, and [SII] 6716+6731 \AA{}, can be used  to determine the main ionising source at each position. The lines' intensities are measured after subtraction of the continuum, exploiting the pure stellar emission best fit model provided by {\sc sinopsis}, to take into account any possible contamination from stellar photospheric absorption. Only spaxels with a $S/N> 3$ in all the emission lines involved are considered.
All the diagnostic diagrams \citep[BPT,][]{Baldwin1981}  are concordant in finding that  young stars produce the ionised gas \citep[``Star-forming'' according to][]{Kauffmann2003, Kewley2006} throughout the galaxy and in excluding the presence of AGN in the galaxy center (see the [OIII]/\Hb vs [NII]/\Ha plot in the left panel of Fig.\ref{fig:BPT}, the other plots are not shown). 
This is in agreement with previous classifications found in the literature for the same galaxy \citep[e.g.,][]{Veron2010}.

Since the ionisation source  is mostly photoionisation by young stars, we can now measure the total ongoing SFR, obtained from the dust- and absorption-corrected \Ha luminosity   adopting the \cite{Kennicutt1998a}'s relation for a \cite{Chabrier2003} IMF.
Integrating the spectrum over the galaxy, we get a value of SFR=0.74 $\rm{M_\odot \, yr^{-1}}$. 

Figure \ref{fig:P95080} presents the maps of other quantities obtained from the MUSE datacubes. From left to right, top to bottom it shows the gas and stellar kinematics, the metallicity of the ionised gas,  the extinction map, the luminosity weighted age and the surface mass density. 
Panels (a) and (b)  show the \Ha velocity and  velocity dispersion maps. The gas is rotating around the North-south direction, the East side is receding, the West side is approaching. The velocity field is quite regular and spans the range ($-100<$v/\kms$<100$). The median error on the gas velocity  in  the spaxels is $\sim$5 \kms.  Uncertainties on the stellar motion are the formal errors of the fit calculated using the original noise spectrum datacube and have been normalized by the $\chi^2$ of the fit. The velocity of the detached clouds is that expected given their position with respect to the galaxy, suggesting that they indeed belong to the object. To further assess the values obtained for the velocity of the clouds, we integrated the spaxels of each cloud and run  \kube on the integrated spectra. Values obtained on the spatially resolved and integrated spectra are largely in agreement. The few spaxels in the West region of the galaxy with velocity $\sim$100 \kms are residuals of the sky line emission discussed above and do not carry any information.  

The velocity dispersion is overall low, having a median value of 16 \kms. This is indicative of a dynamically cold medium. The East side has a systematically higher velocity dispersion than the West one.  All the clouds have a typically low velocity dispersion.  

Panels (c) and (d) of Fig. \ref{fig:P95080} show the stellar velocity and stellar  velocity dispersion maps, respectively, for Voronoi bins with S/N> 10. The velocity field of the  stellar component of P95080 is similar to that of the gas, spanning a similar range ($-100<$v/\kms$<100$), though less spatially extended. The median error in stellar velocity is $\sim50$ \kms. The bending of the locus of zero-velocity is due to the presence of the bar{, as discussed in Erroz-Ferrer et al. (2015)}.
Also the velocity dispersion of the  stars is typically low ($<20$ \kms, which is below the resolution limit). Deviations are seen in the eastern part of the galaxy, where larger errors prevent us from drawing solid conclusions.

Panel (e) of Fig. \ref{fig:P95080} presents the spatial distribution of the metallicity of the ionised gas, i.e. $12+\log[O/H]$. Only the spaxels with S/N is >3 for all the lines involved in the computation of the metallicity ([NII]6585/[SII]6717+6731 vs [OIII]/[SII]6717+6731) are plotted. P95080 is characterised by  quite high values of the metallicity in the center ($12+\log[O/H]\sim 9$) and then by a smooth decline towards the outskirts, which are characterised by  $12+\log[O/H]\sim 8$. P95080 lays on the typical mass-metallicity relation  for local  field galaxies  \citep[see][]{Tremonti2004}.
Unfortunately, given the low S/N of some of the lines, we can not properly constrain the metallicity for the clouds. 

Panel (f) shows the $A_V$ maps for spaxels with a S/N(\Ha)> 3.  Overall, P95080 is characterised by low values of extinction, almost always $<1$mag. Relatively higher values of extinction are found preferentially in the central regions and trace one of the spiral arms of the galaxy.

Panel (g) presents the map of the luminosity weighted ages. This provides an estimate of the average age of the stars weighted by the light we actually observe, and gives an indication on when the last episode of star formation occurred. The map  shows that  in the central regions the typical luminosity weighted age of the galaxy is $\sim 10^{9-9.5}$ yr, and it decreases towards the outskirts, where it reaches values of $\sim10^{7}$ yr. While in the center the distribution of ages is quite homogeneous, towards the outskirts it becomes more patchy, showing many knots of younger ages. Typically, they corresponds to the \Ha blobs seen in the top right panel of Fig. \ref{fig:rgb_image}. 

Finally, panel (h) shows the stellar mass density. The vast majority of the mass is confined in the central parts of the galaxy, while the outskirts, and especially the clouds around the galaxy, are extremely less massive, reaching a mass density of  $\rm 3\times 10^6 M_\odot/kpc^2$. The bar and two main spiral arms, already detected in the \Ha map, are seen also in the stellar mass density. Running \sinopsis on the integrated spectra of the entire galaxy, we obtain a total M$_\ast$ of 9.5$\rm \times 10^{9} \, M_\odot$.

Given its values of SFR and stellar mass, P95080  lays on the typical SFR-mass relation for star-forming field galaxies  \citep[Paper XIV]{Vulcani2018c}.

\subsection{P19482}

\begin{figure*}
\centering
\includegraphics[scale=0.51,clip, trim=0 0 0 0]{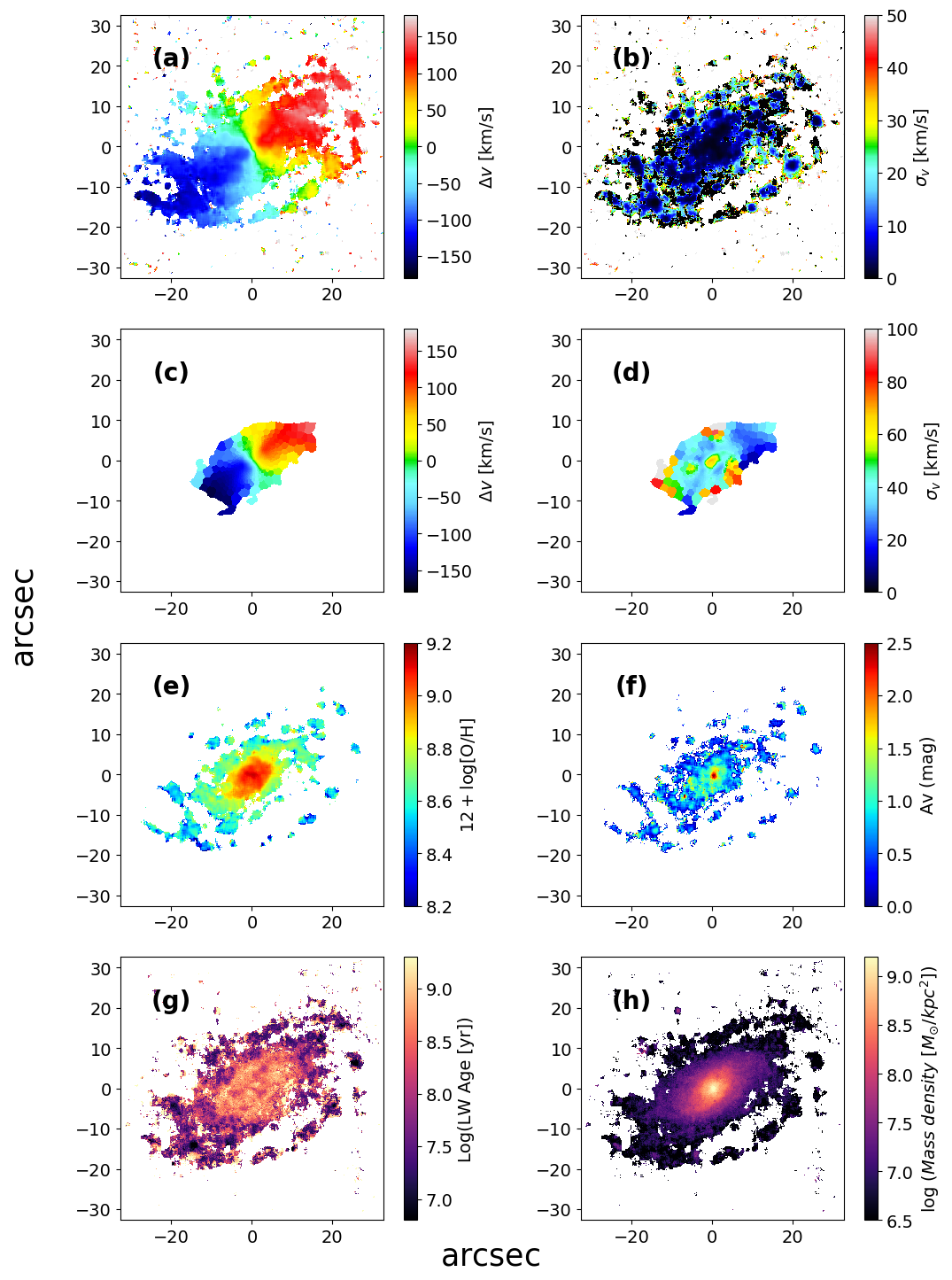}
\caption{P19482. Panels are as in Fig.\ref{fig:P95080}.  \label{fig:P19482} }
\end{figure*}

{ We now focus on P19482, whose color composite image is presented in the second row of Fig. \ref{fig:rgb_image}. This is a spiral galaxy, with a slightly higher inclination than P95080.  
A spiral  arm extends towards South-West. {  We detect  the presence of 42 clouds}, whose size is much larger than the typical size of the noise, seen e.g. in the corners of the image. Most of the  detached clouds follow the spiral arms, but especially in the North -East region no stellar disk seems to be associated to the presence of the clouds. The maximum extension of the \Ha disk is $7.9\times r_e$.

Figure \ref{fig:P19482} present all the other properties of the galaxy. The analysis of the velocity field (panel (a)) indicates that these clouds belong to the galaxy, as their velocity is consistent with that of the part of the galaxy that is close to them. As we did for P95080,  we integrated the spaxels of each cloud and run  \kube on the integrated spectra. Values obtained on the spatially resolved and integrated spectra are largely in agreement.

Overall, in each position, the gas and the stars (panel (c)) rotate around the same axis and at similar speed ($-180<$v/\kms$<180$). 
The median error on the gas velocity is 4 \kms, the one on the stellar velocity is 50 \kms. 

The gas velocity dispersion (panel (b)) is overall $<10$ \kms. The  median error on the gas velocity dispersion is $\sim 6$ \kms. 

The velocity dispersion of the stellar component (panel (d)) is overall quite low ($\sim40$ \kms). In the South East region it is systematically higher, reaching values of 80 \kms.

The analysis of the diagnostic diagrams (central panel of Fig.\ref{fig:BPT}) does not detect the presence of an AGN
 in the galaxy center. The emission-line ratios are  consistent with gas being photoionised by young stars.

The metallicity map of the ionised gas, presented in panel (e), ranges from $12+\log[O/H]=8.5$ in the outskirts to $12+\log[O/H]=9.5$ in the core. This value sharply decline towards the outskirts, where the typical metallicity is $\sim 8.5$. We can estimate the metallicity also in few clouds, finding that this is consistent with the edges of the galaxy. 

P19482 presents low values of dust extinction (panel (f)) ranging from 0.2 mag in the outskirts and 2 mag in the center. 

The luminosity weighted age (panel (g)) strongly varies across the galaxy. Maximum values are reached in the galaxy center, where LWA$\sim 10^9$ yr, while moving towards the outskirts, the luminosity weighted age progressively decreases, down to a minimum vlaue of LWA$\sim 10^7$ yr. 
As for P95080, most of the mass (panel (h)) is contained in the central part of the galaxy. 
Running \sinopsis on the integrated spectra of the entire galaxy, we obtain a total M$_\ast$ of 1.9$\rm \times 10^{10} \, M_\odot$.

The total ongoing SFR of P19482 is 1.3 $\rm{M_\odot \, yr^{-1}}$, its position on the SFR-mass relation is that expected for typical for star-forming field galaxies  \citepalias{Vulcani2018c}. 

}
\subsection{P63661}

We now move our attention to P63661, whose RGB image is presented in the third row of Fig. \ref{fig:rgb_image}. Similarly to the previous galaxies, this is a spiral galaxy, with a moderate inclination.  
A spiral unwinding arm extends towards South-West. 

\begin{figure*}
\centering
\includegraphics[scale=0.51,clip, trim=0 0 0 0]{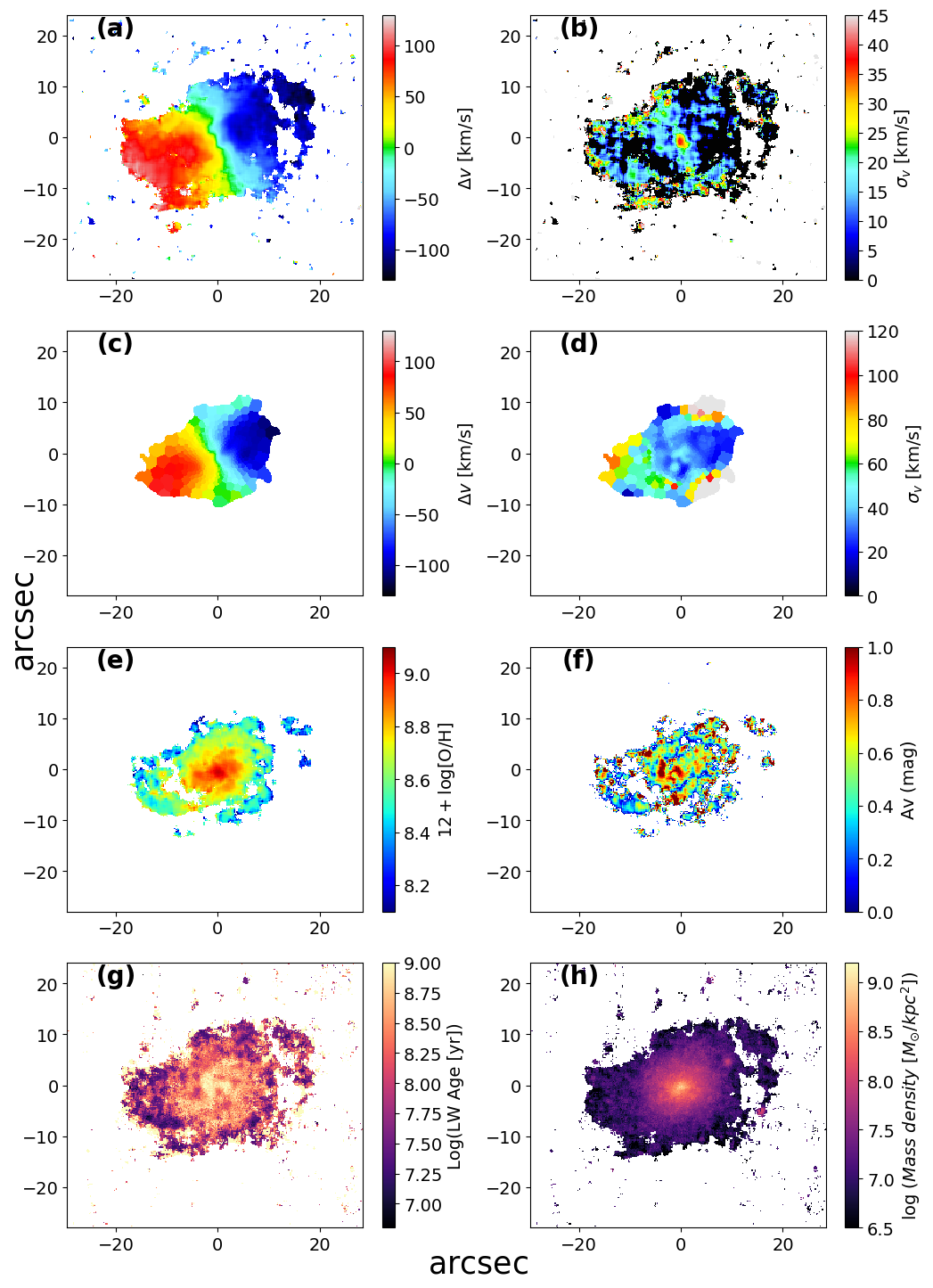}
\caption{P63661. Panels are as in Fig.\ref{fig:P95080}.  \label{fig:P63661} }
\end{figure*}

{ The \Ha map shown in  Figure \ref{fig:rgb_image} and Figure \ref{fig:P63661} unveils a much more complicated structure. 
Focusing on 
the \Ha distribution, we find that it extends well beyond the stellar disk. The maximum extension of the \Ha disk is $4.3\times r_e$}. On the West side of the galaxy, the \Ha map presents a rift. A portion of the gas is detached from the main body. We remind the reader that GASP data reach a surface brightness detection limit of $V\sim 27$ mag arcsec$^{-2}$ and $\rm{\log (H\alpha [erg \, s^{-1} \, cm^{-2} arcsec^{-2}]) \sim -17.6}$ at the 3$\sigma$ confidence level \citepalias{Poggianti2017a}. 

On the North-East side of the galaxy, the \Ha distribution is somehow broken off and presents a sharp edge. 

In addition, a number of smaller clouds surround the galaxy {  for a total of 16.}
. From the analysis of the velocity field (panel (a)) it results that these clouds belong to the galaxy, as their velocity is consistent with that of the part of the galaxy that is close to them. Running \kube on the integrated spectra of  each cloud, we obtained values compatible with those obtained on the  spaxels. 

Overall, in each position, the gas and the stars (panel (c)) rotate around the same axis and at similar speed ($-100<$v/\kms$<100$). In the external regions, where there are no stars in correspondence of the gas, the gas reaches velocities of $|v|\sim 120$ \kms. 
The median error on the gas velocity is 4 \kms, the one on the stellar velocity is 40 \kms.

The gas velocity dispersion (panel (b)) is overall $<20$ \kms, except in the core, where it reaches values of $\sim 45$ \kms. The  median error on the gas velocity dispersion is <2 \kms. 

The velocity dispersion of the stellar component (panel (d)) is more chaotic, especially towards South-East, where a spiral arm is present. Nonetheless, typical values do not exceed $\sim$50 \kms.

No central AGN is detected from the analysis of the diagnostic diagrams (central panel of Fig.\ref{fig:BPT}). The emission-line ratios are  consistent with gas being photoionised by young stars. This finding confirms previous classifications  \citep[e.g.,][]{Veron2010}.

The metallicity map of the ionised gas is presented in panel (e). The central part of the galaxy has a $12+\log[O/H]>9$. This value sharply decline towards the outskirts, where the typical metallicity is $\sim 8.5$. From what we can infer from the significantly meaningful spaxels in the detached part of the galaxy towards West, the metallicity of the region is significantly lower. 

The $A_V$ map (panel (f)) shows that overall P63661  is characterised by low values of extinction, almost always $\leq1$mag. 

The last two panels of Fig.\ref{fig:P63661} show the properties of the stellar populations. The maximum luminosity weighted age of the galaxy (panel (g)) is found in the galaxy center: LWA$\sim 10^9$ yr. Moving towards the outskirts, the luminosity weighted age constantly decreases, to reach the minimum values in the detached region in the North-West side and towards East, along the extension of one spiral arm. Most of the mass (panel (h)) is contained in the central part of the galaxy. Beyond the R25 the stellar mass density reaches values of  $\rm 3\times 10^6 M_\odot/kpc^2$. Running \sinopsis on the integrated spectra of the entire galaxy, we obtain a total M$_\ast$ of 1.8$\rm \times 10^{10} \, M_\odot$.

The total ongoing SFR of P63661 is 0.86 $\rm{M_\odot \, yr^{-1}}$, its position on the SFR-mass relation is that expected for typical for star-forming field galaxies  \citepalias{Vulcani2018c}.

\subsection{P8721}

P8721 is a spiral galaxy with a quite high inclination. Projection effects can therefore affect the interpretation of the results and possible detached clouds seen in projection might appear as part of the galaxy. 

\begin{figure*}
\centering
\includegraphics[scale=0.51,clip, trim=0 0 0 0]{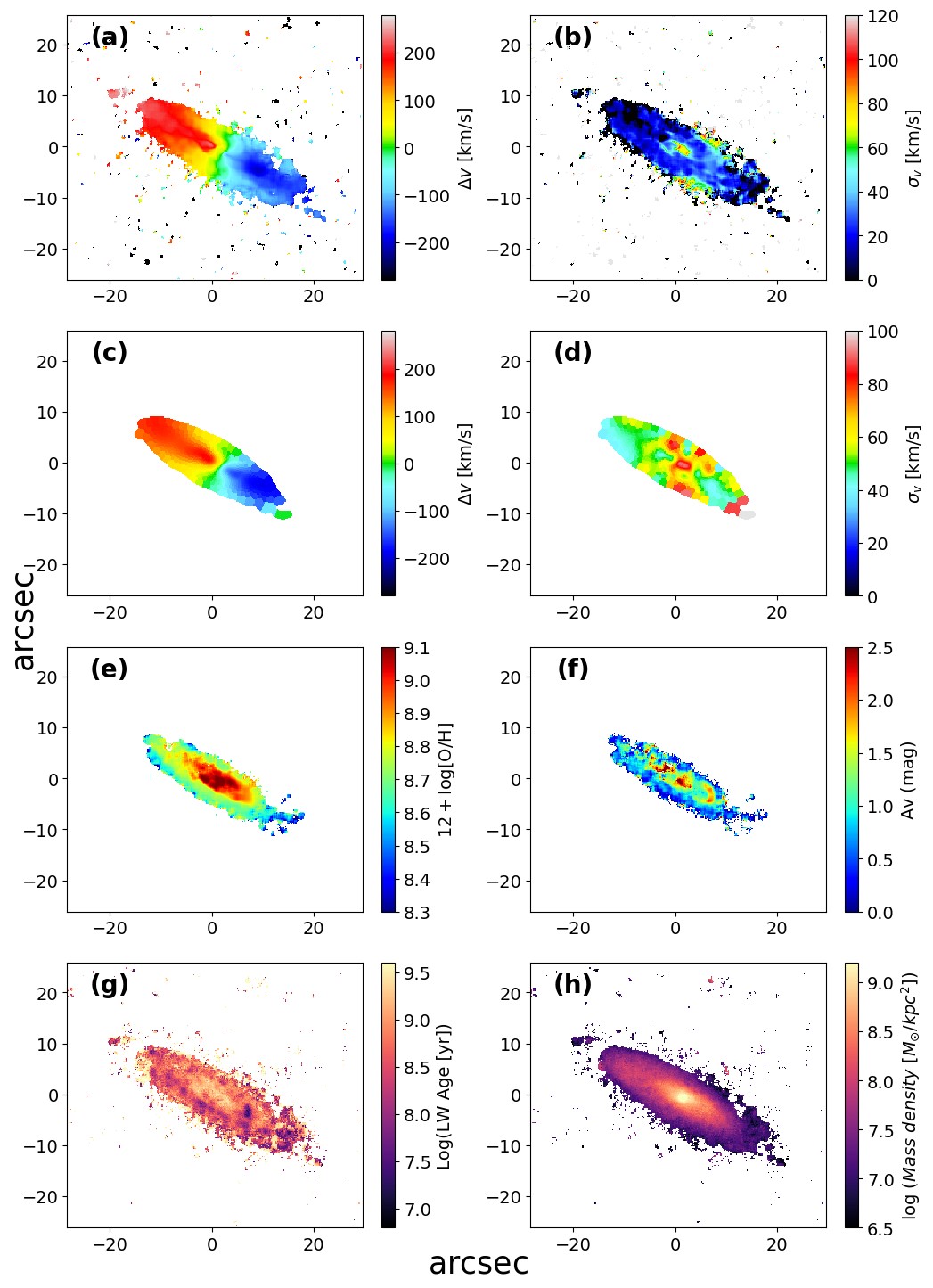}
\caption{P8721. Panels are as in Fig.\ref{fig:P95080}. \label{fig:P8721} }
\end{figure*}
The ionised gas is much  extended, especially towards South-West. The maximum extension of the \Ha disk is $4.6\times r_e$.
Strikingly, Fig.
\ref{fig:rgb_image} shows that the stellar disk extends mostly towards North-East with respect to the galaxy center while the ionised gas disk extends mostly towards South-west. It therefore appears that the light distribution in B and \Ha are distinct. A bow of bright \Ha knots is visible in the Southern part of the galaxy. 
{  12} clouds of detached gas are visible both towards South-West and towards North-East. The velocity of the gas (panel (a) in Fig. \ref{fig:P8721}) in these clouds is consistent with them belonging to P8721. 

The gas velocity field is regular and spans the range -220$<v/$\kms$<$220.  The  median error on the gas velocity  is $\sim4$ \kms. 
The gas velocity dispersion, shown in panel (b) is overall quite low ($<35$ \kms), except for the core and two external regions. The median error on the gas velocity dispersion is $<4$ \kms. The right panel of Fig.\ref{fig:BPT} shows that while an AGN is not detected, the central region of the galaxy has a composite spectrum, indicative of either shocks or old evolved stars.

The stellar kinematics (panels (c) and (d)) is  regular, except for a protuberance in the southern region, probably due to a spiral arm,  and similar to that of the gas in the same spatial position. The  median error on the stellar velocity  is $\sim25$ \kms.  The stellar velocity dispersion ranges from 40 \kms to 80 \kms. 
The metallicity of the ionised gas (panel (e)) is  $12+\log[O/H]\sim 9.1$ within R25 and then abruptly decreases. It reaches minimum values in the tail towards South West. The two sides of the galaxy (SW and NE) show different slopes of the gradients.  No metallicity values are reliable for the gas in the clouds. 

The $A_V$ map (panel (f)) shows a peak of dust attenuation in the core of the galaxy ($A_V\sim 2.5$ mag) and a decline towards the outskirts, where it reaches values of $\sim 0.5$ mag.

\begin{figure*}
\centering
\includegraphics[scale=0.3]{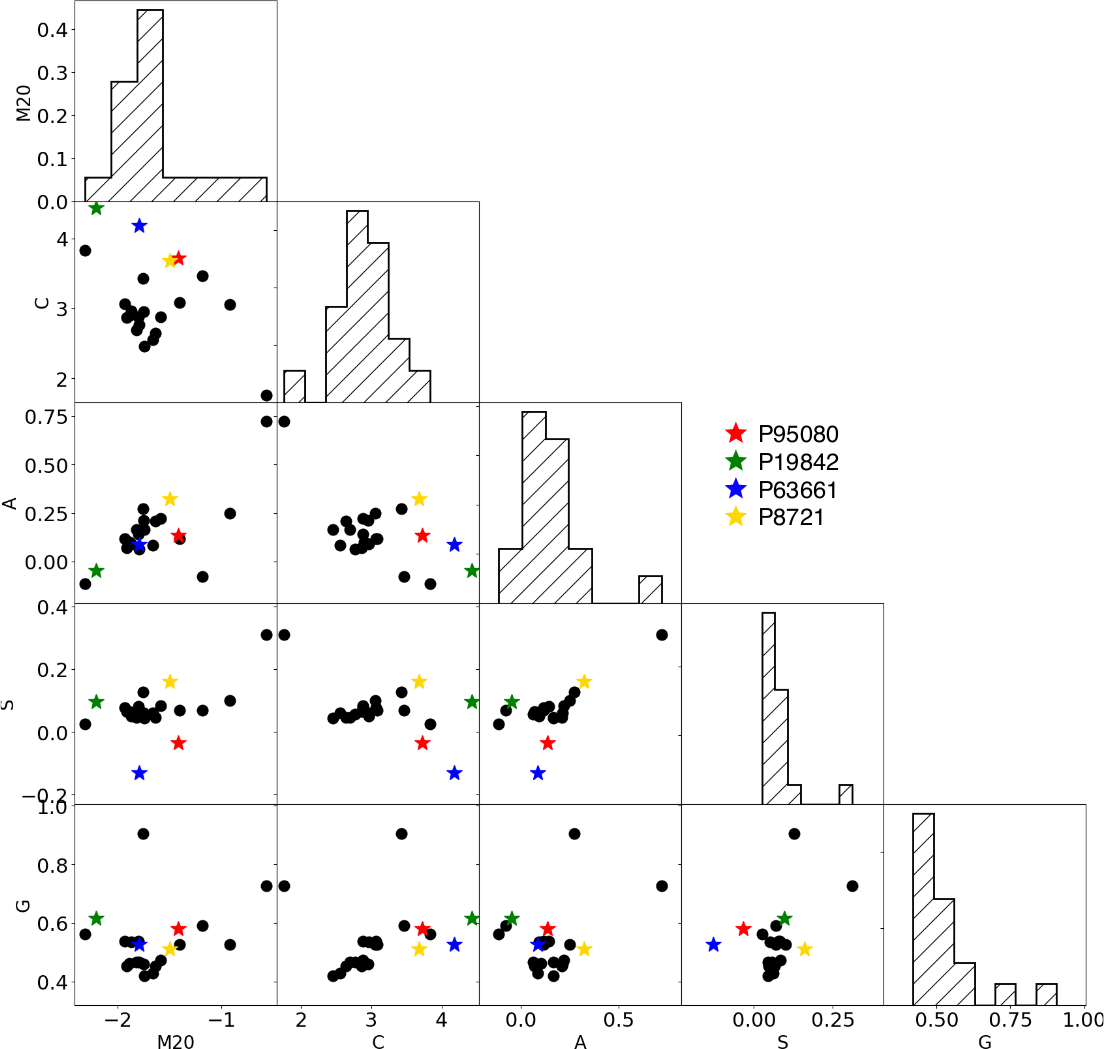}
\includegraphics[scale=0.3]{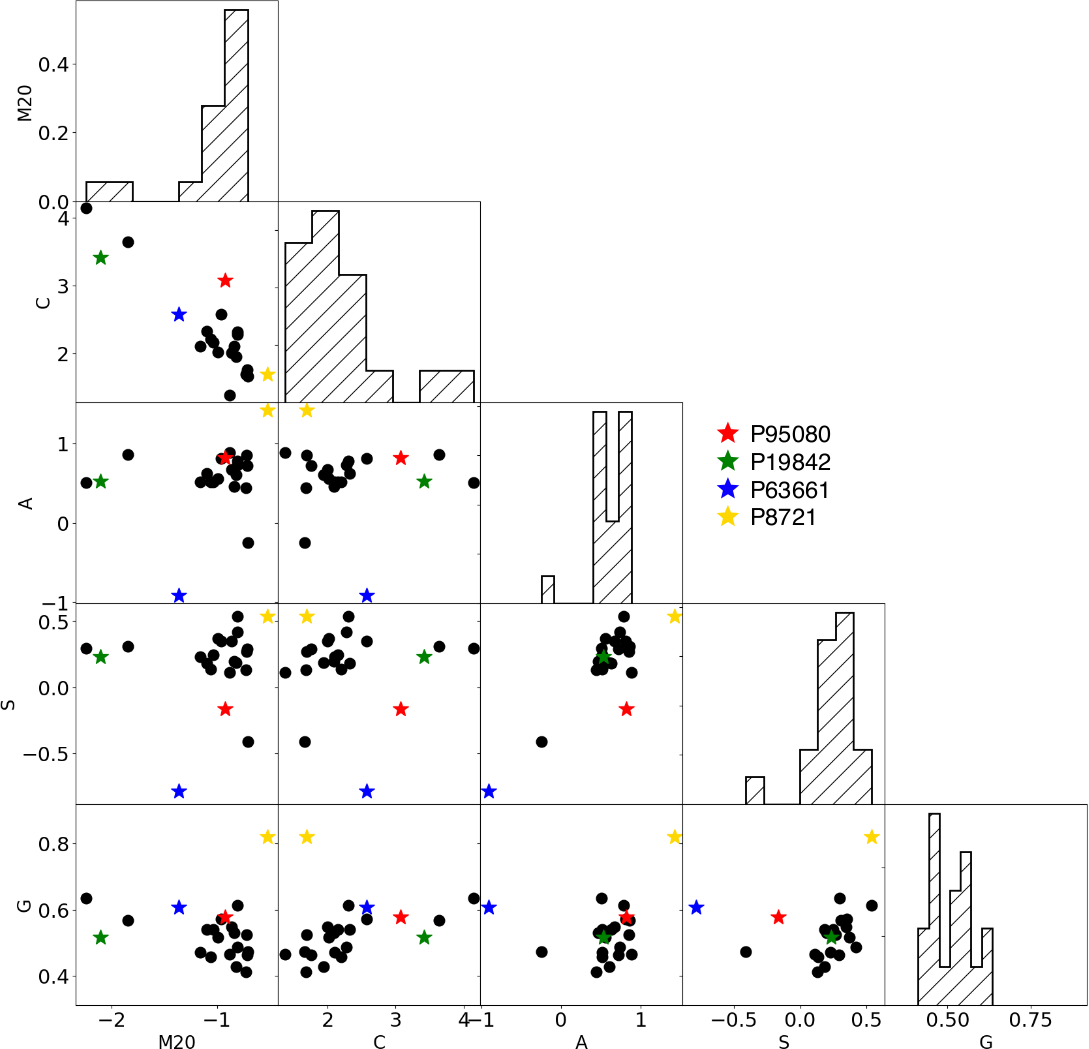}
\caption{The morphological parameters G, M20, C, A and S for the three galaxies presented in this paper (colored stars) and galaxies of a GASP control sample visually selected for not having morphological distortions \citepalias{Vulcani2018c}(black points). The histograms show the distribution of the parameters.  Left panels are based on the continuum underlying \Ha, right panels on the \Ha images. \label{fig:morphs} }
\end{figure*}

The luminosity weighted age (panel (g)) is slightly older within R25, with typical values around 10$^{9.5}$ yr. Outside the R25, it has average values around 10$^{8.5}$ yr.
The youngest region of the galaxy is found in the South-West part of the object. 

The mass density map (panel (h)) shows that the bulk of the mass in located in the galaxy core ($\rm  10^9 M_\odot/kpc^2$), while the South-West part of the object gives a very little contribution to the total mass of P8721.
Running \sinopsis on the integrated spectra of the entire galaxy, we obtain a total M$_\ast$ of 5.6$\rm \times 10^{10} \, M_\odot$.

The total ongoing SFR  is 1.05 $\rm{M_\odot \, yr^{-1}}$, it therefore lays on the SFR-mass relation of  star-forming galaxies in the field \citepalias{Vulcani2018c}.

\subsection{Morphological analysis}\label{sec:morph}

To further assess the peculiarity of the light distribution in these galaxies, we run a number of non-parametric morphological measurements, exploiting the python package statmorph \citep{Rodriguez2018}, on the images of the continuum underlying the \Ha (red continuum) and the \Ha images, obtained from the fits of \kube. Specifically, we measured:
\begin{itemize}

\item {\it Concentration C}: Ratio of the circular radius containing 80 per cent ($r_{80}$) of a galaxy's light to the radius containing 20 per cent ($r_{20}$) of the light \citep{Bershady2000, Conselice2003, Peth2016}. A large concentration value indicates a majority of light is concentrated at the center of the galaxy, i.e the presence of a bulge.
\item {\it Asymmetry A}: Difference between the image of a galaxy and the galaxy rotated by 180 degrees \citep{Conselice2000, Peth2016}. This determines a ratio of the amount of light distributed symmetrically to all light from the galaxy. A large value of asymmetry indicates that most of the light is not distributed symmetrically. 
\item {\it Gini Coefficient G}: Measure of the equality of light distribution in a galaxy \citep{Lorenz1905, Abraham2003,  Lotz2004, Conselice2014}. A value of G = 1 is obtained when all of the flux is concentrated in a single pixel,a value of G = 0  when the brightness distribution is homogeneous. 
\item {\it M20}: Second order moment of the brightest regions of a galaxy \citep{Lotz2004} tracing the spatial distribution of any bright clumps.  It is sensitive to bright structure away from the center of the galaxy; flux is weighted in favor of the outer parts. It therefore is relatively sensitive to merger signals and tidal structures, specifically star-forming regions formed in the outer spiral or tidal arms. If no such structures are in the image, the 20\% brightest pixels will most likely be concentrated in the center of the galaxy, which is weighted lower. Low values of M20 are obtained for smooth galaxies with bright nucleus (Ellipticals, S0 or Sa),  much higher values (less negative) for galaxies with extended arms featuring bright H{\sc ii}  regions. 
\item {\it Smoothness S}: Degree of small-scale structure \citep{Conselice2003, Takamiya1999}. Larger values of S actually correspond to galaxies that are less smooth (i.e. more `clumpy'). 
\end{itemize}

Figure \ref{fig:morphs} compares the values of the morphological measurements of the three galaxies under inspection to those of a GASP (field+cluster) control sample carefully selected for not having morphological distortions \citepalias{Vulcani2018c}. We carefully checked that all the galaxies of the control sample do not have any similar detached cloud. 
As far as the red continuum is concerned, the { four} galaxies present relatively high concentration  values, slightly higher than the bulk of the control sample.  They also present similar asymmetry and Gini values, indicating that stars are symmetrically and quite homogeneously distributed. { Only P19482 is offset, indicating the presence of dishomogeneites also in the stellar component.}
They do not stand out in the M20-Asymmetry and M20-Gini planes, excluding ongoing mergers for these objects \citep[e.g.][]{Lotz2008a, Lotz2008b}. 
Considering \Ha, it emerges that the star formation is less concentrated than the stars, both for these galaxies and the control sample, P8721 is one of the less concentrated objects of all galaxies. Moving to asymmetry and smoothness, overall all galaxies are characterised by higher absolute values than for the stellar continuum. P8721 and P63661 really stand out in these distributions. P8721 is peculiar also in terms of Gini coefficient, while the other two galaxies follow the control sample trends. { P19482 present low values of M20, confirming the presence of many clumpy star forming regions spread across the disk.}

Taken together, these results suggest that as for the red continuum, which traces the stars, galaxies are ``normal''. 
Peculiarities with respect to a control sample of undisturbed galaxies emerge when looking at the star forming regions only. 
 The \Ha distribution is clumpy and with small scale structures. P8721 is the most peculiar object, followed by P63661 and P19482, while P95080 is  more regular. 

This analysis therefore corroborates the previous analysis based on visual morphology that all the three galaxies are regular when the stellar light distribution is considered. In almost all the cases, the evidence for anomalies is stronger when the \Ha images are analysed, suggesting that the ionised gas is the most disturbed component.

\section{Discussion}

In the previous section we have described the spatially resolved properties of three galaxies that present peculiar light distributions and a number of common features. 
First of all, they are all characterised by a ``tattered'' \Ha distribution. They all present \Ha clouds  beyond  the stellar disk, up to several kpc. They are visible even in P8721, an unfavored case given its high inclination. 
These clouds have typically a size of 3-5 kpc, but a detached region $\sim 20$ kpc long is visible in P63661.  Within the GASP sample, these, along with P5215 discussed in \citetalias{Vulcani2018b}, are the only galaxies presenting such extended and peculiarly tattered gas distribution. 

According to the gas kinematics, these clouds do belong to the galaxy: their velocity is similar to that of the closest part of the galaxy. 
Both the gas and the stellar kinematics are regular and  resemble each other. We can therefore exclude processes that involve a redistribution of the stellar orbits, such as mergers \citepalias[see, e.g.,][]{Vulcani2017c} or processes strongly affecting the gas distribution, such as strong ram pressure stripping (see, e.g., \citealt{Gunn1972}, \citetalias{Poggianti2017a}).
{ Simulations by e.g. Jesseit et al. (2007), Kronberger et al. (2007)  have indeed studied  the  2D kinematic analysis for a sample of simulated binary disc merger remnants with different  mass ratios, showing how merger remnants usually show a multitude of phenomena, such as  heavily distorted velocity fields, misaligned rotation, embedded discs, gas rings, counter-rotating cores and kinematic misaligned discs. None of these features are evident from our maps.}

At least other two pieces of evidence exclude that the galaxies have undergone  a recent merger. On one side, the BPT diagrams  presented in Fig. \ref{fig:BPT} show { that there are no signs of tidally induced shocks, associated with the interaction process, contributing to the ionization of the gas (Colina, Arribas \& Monreal-Ibero 2005).  Extended shock ionization has been previously reported in local U/LIRGs (Monreal-Ibero, Arribas \& Colina 2006; Rich et al. 2011; Rich, Kewley \& Dopita 2014). In all these cases, shock ionization exhibits characteristics of extended low-ionization nuclear emission-line region (LINER)-like emission with broadened line profiles. In  P95080, P19482 and P63661 we find no broadened line profiles falling in the so-called composite region. Only P8721 has a few central spaxels characterised by  composite spectra, but composite regions due to interactions would be expected more in the outer parts of the galaxy. }

In addition, we see continuous distribution, not different sequences, confirming again that we are  observing gas belonging to  one galaxy. Indeed, in cases of e.g. mergers or gas accretion we should see different line ratios indicative of different chemical abundances in the different regions of the galaxies { \citepalias[see, e.g., Fig. 12 in][]{Vulcani2018b}}. 
On the other side, the observed metallicity distribution has a smooth gradient, suggesting that no strong process altered it considerably. 

The asymmetry of the metallicity gradient in P8721 might however suggest that this galaxy is accreting low metallicity gas from the Southern side, similarly to what presented for another GASP galaxy in \citetalias{Vulcani2018}, but no other pieces of evidence support this scenario. P8721 is also the most peculiar object when the morphological analysis on the \Ha is performed, being at the tail of the distributions in all the quantities analysed.

\begin{figure*}
\centering
\includegraphics[scale=0.27]{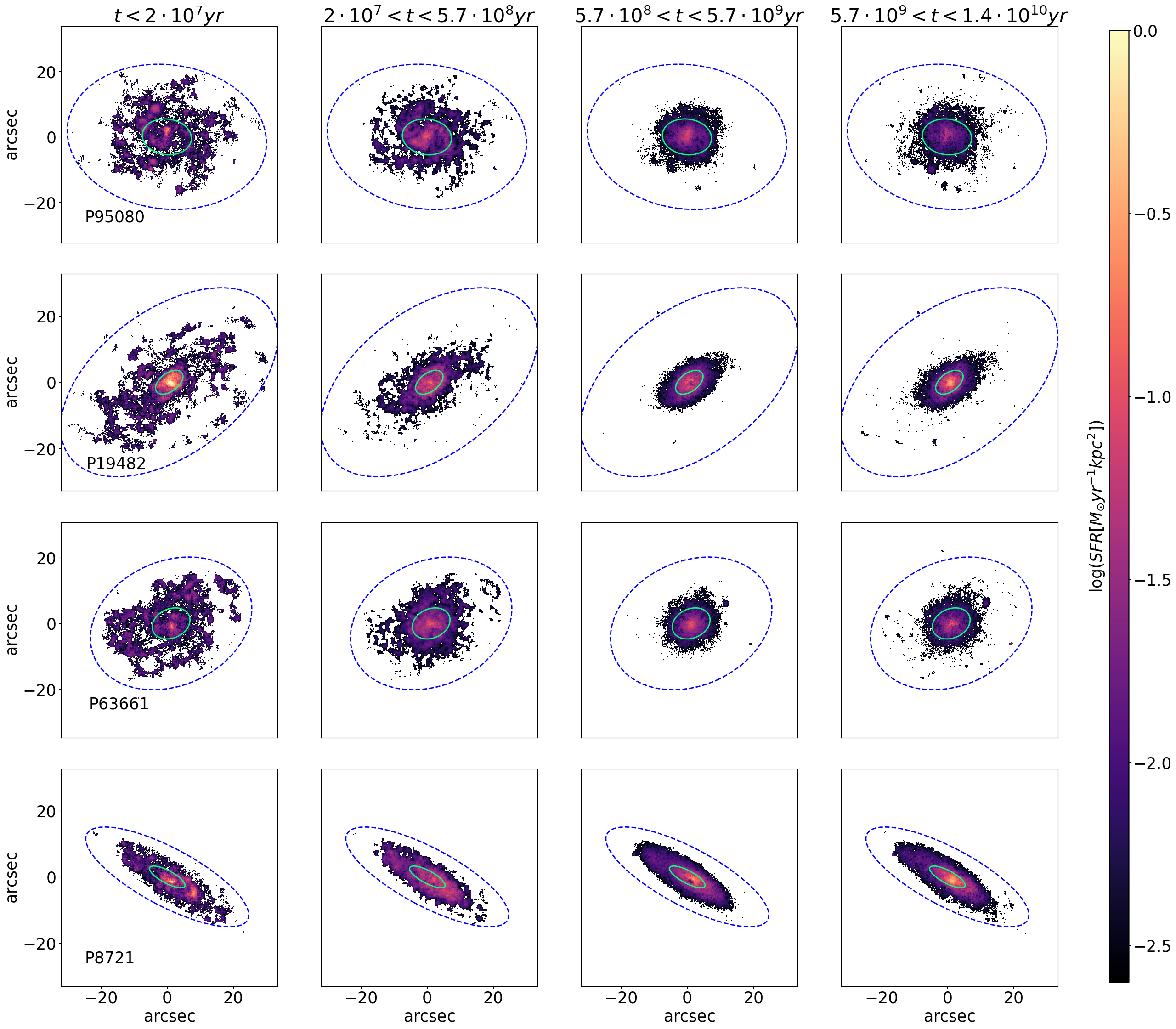}
\caption{Stellar maps of different ages, illustrating the average star formation rate per kpc$^2$ during the last $2\times 10^7$ yr (left), between $2\times 10^7$yr and $5.7 \times 10^8$yr (central left), $5.7 \times 10^8$yr and $5.7 \times10^9$yr (central right) and $> 5.7 \times 10^9$yr ago (right), for P95080 (upper), P63661 (central) P8721 (bottom). In all the plots, the green ellipses show the  $r_e$, the dashed blue ellipses show the maximum radius at which \Ha is detected (see text for details). 
\label{fig:SFH} }
\end{figure*}

\begin{figure*}
\centering
\includegraphics[scale=0.25]{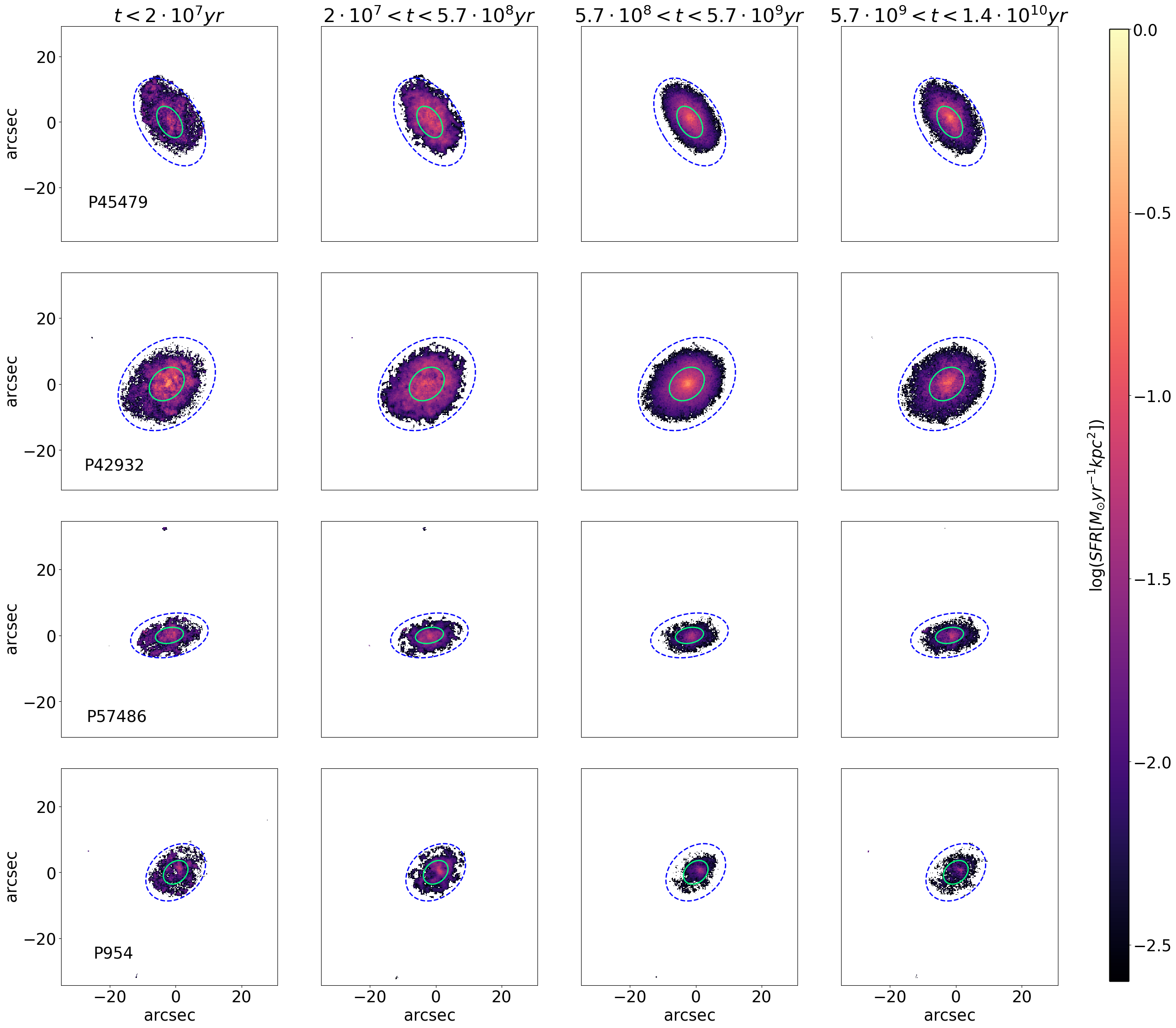}
\caption{Same as \ref{fig:SFH}, but for galaxies in the control sample. 
\label{fig:SFH_control} }
\end{figure*}

In addition to looking at the luminosity weighted age maps, to better investigate the mode of growth of these galaxies (inside out or inside in) in Fig. \ref{fig:SFH} we present the galaxy spatially resolved star formation histories. 
These plots show the variation of the  SFR  across cosmic time in  four age bins in such a way that the differences between the spectral characteristics of the stellar populations are maximal (\citealt{Fritz2007} and \citetalias{Fritz2017}). Note that {\sc sinopsis} tends to include an unnecessary small percentage of old (t>$5.7\times 10^8$) stars when the spectra have a low signal-to-noise values. To be conservative, we neglect the contribution of stars older than $5.7\times 10^8$ yr in low S/N spectra (S/N$<$3). The entire  contribution of young stars, instead, is taken into account, given the fact that it is estimated from the emission lines, which are more reliable features. 

In Fig.  \ref{fig:SFH} each row corresponds to a galaxy, each column to a different age bin. For P95080, { P19482} and P63661 the inside-out growth is outstanding. In the two oldest age bins ($t>5.7\times 10^8$ yr ago), the star formation mostly occurred within half of the current { \Ha maximum disk} and only in recent epochs ($t<5.7\times 10^8$ yr ago) the outer part of the disk started to form.
The maximum spatial extension of the star forming disk is observed in the current age bin ($t<2\times 10^7$ yr).  In P8721, instead, the SFR is relatively constant with time overall in the galaxy and { external regions} might have been already forming stars even in the oldest age bin. This is another piece of evidence that distinguishes P8721 from the other galaxies.

{ For comparison, Figure \ref{fig:SFH_control} shows the maps of SFR in the oldest and in the youngest age bins for the four galaxies in the control sample already presented in Fig.\ref{fig:rgb_image_control}. In these galaxies the spatial extension of the disk is very similar at the two ages, suggesting that not all galaxies are characterised by strong inside-out growth.}

The next step to better understand  the possible physical mechanisms acting on these galaxies is to characterise the environment around them.

\subsection{The environments}
To characterise the environments of the galaxies we have presented in the previous section, we exploit two publicly available catalogs. Both are based on the spectroscopic sample of the galaxies of SDSS data release 10, complete down to m$_r$= 17.77 mag. The first catalogue identifies galaxy groups and clusters and was published by \cite{Tempel2014_g}. The  second catalogue identifies galaxy filaments and was published by \cite{Tempel2014_f}.

\begin{table*}
\caption{Properties of the groups hosting  the galaxies. Values are taken from \citet{Tempel2014_g}. For each group, the redshift (z$_{\rm gr}$), the coordinates (RA$_{\rm gr}$ and DEC$_{\rm gr}$), the number of group members (N$_{\rm gals, \, gr}$), the virial radius R$_{\rm vir, \, gr}$, the mass of the halo both assuming a NFW and a Hernquist profile ($\log M^{NFW}_{\rm halo, \, gr}$, $\log M^{Her}_{\rm halo, \, gr}$) are given. P63661b is the bigger cluster close to the system of P63661.  \label{tab:groups}}
\centering
\setlength{\tabcolsep}{2pt}
\begin{tabular}{crrrrrrrrrrrrr}
\hline
  \multicolumn{1}{c}{ID} &
  \multicolumn{1}{c}{z$_{\rm gr}$} &
  \multicolumn{1}{c}{RA$_{\rm gr}$} &
  \multicolumn{1}{c}{DEC$_{\rm gr}$} &
  \multicolumn{1}{c}{N$_{\rm gals, \, gr}$} &
  \multicolumn{1}{c}{R$_{\rm vir, \, gr}$}    &
  \multicolumn{1}{c}{$\log M^{NFW}_{\rm halo, \, gr}$} & 
  \multicolumn{1}{c}{$\log M^{Her}_{\rm halo, \, gr}$}   \\
  \multicolumn{1}{c}{} &
  \multicolumn{1}{c}{} &
  \multicolumn{1}{c}{(J2000)} &
  \multicolumn{1}{c}{(J2000)} &
  \multicolumn{1}{c}{} &
  \multicolumn{1}{c}{(kpc)} &
  \multicolumn{1}{c}{(M$\odot$)} &
   \multicolumn{1}{c}{(M$\odot$)}   \\
\hline
P95080	&0.04136	&198.08969	&-0.23002	&3	 & 315 & 12.63 & 12.84 \\
P19482	&-	&-	&-	&-	 & - & - & - \\
P63661	&0.05597	&218.06171	&0.17165	&2	 & 232 &12.58 &12.58 \\
P63661b	&0.05612	&217.49573	&0.30415	&32	 & 493 &13.77 &13.99\\
P8721	&0.06609	&158.51371	&0.00926	&4	 & 276 & 11.61 & 11.81\\
\hline
\end{tabular} 
\end{table*}

\begin{figure*}
\centering
\includegraphics[scale=0.36,clip, trim=0 0 0 0]{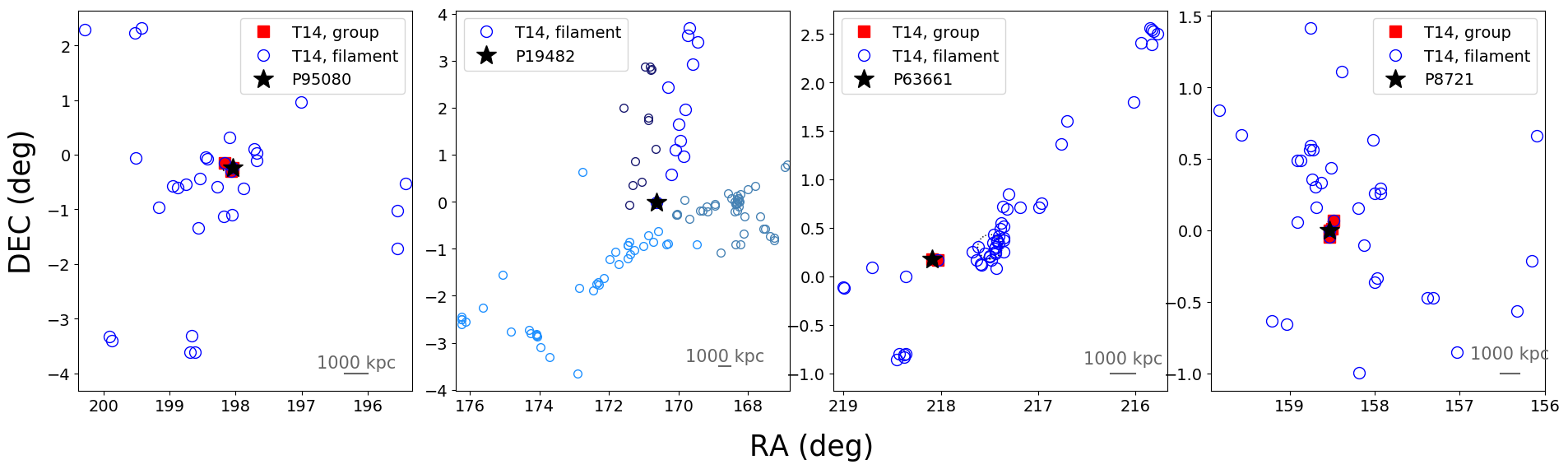}
\caption{Position on the sky of the targets, represented by the stars. Filled red squares represent galaxies in groups, according to \citet[][T14, group]{Tempel2014_g}. Empty circles represent galaxies in filaments, according to \citet[][T14, filament]{Tempel2014_f}.  Dashed circles indicate the virial radius of the groups. The scale in the bottom right corner shows 1 Mpc at the redshift of each target. { For P19482, the smaller points with different shades of blue show filaments intersecting the one hosting the galaxy.} \label{fig:envs} }
\end{figure*}

Table \ref{tab:groups} presents some useful information regarding the groups, Fig. 
\ref{fig:envs} shows the position on the sky of the targets and their surroundings. All the values are drawn from \cite{Tempel2014_g}. Besides detecting the filaments, \cite{Tempel2014_f} do not give any quantity useful to better characterise these structures. 

None of these galaxies is located in massive clusters, and { three of them} are  members of small (Milky-Way-like, or slightly more massive, with two or three bright members) groups that are embedded in filaments, { while P19482 most likely does not have any close companion, but is still embedded in a filament}. 

P95080 is part of a three-member group that is located in the center of a long filament { of 32 galaxies}. The galaxy is at 0.5 R$_{\rm vir, \, gr}$ and its closest galaxy is at $\sim$200 kpc (see Tab. \ref{tab:gal_groups}). 

{ P19482 is at the edges of a filament of 12 members and at the intersection among four different filaments all located at the same redshift (z$\sim$0.040-0.044). In total, the structures have more than 100 members. We remind the reader that the identification of filaments is a delicate task (see also the Introduction), therefore it might be that all these galaxies actually belong to the same structure.}
\begin{table}
\caption{Distances of the galaxies from the center of their group, in unit of R$_{\rm vir, \, gr}$ and distance of the closest galaxy, in kpc. For P63661, the value in parenthesis gives the distance from the larger group.  \label{tab:gal_groups}}
\centering
\setlength{\tabcolsep}{2pt}
\begin{tabular}{crr}
\hline
  \multicolumn{1}{c}{ID} &
  \multicolumn{1}{c}{d$_{\rm r_{200}}$} &
  \multicolumn{1}{c}{d$_{\rm closest}$}  \\
  \multicolumn{1}{c}{} &
  \multicolumn{1}{c}{} &
  \multicolumn{1}{c}{(kpc)} \\
\hline
P95080	&0.50 & 193\\
P63661	&0.23 (4.84) & 233\\
P19482 & - & 1750 \\
P8721	& 0.39 & 165\\
\hline
\end{tabular} 
\end{table}

P63661 is part of a binary system and its companion is at a distance of 233 kpc. It is at $\sim 0.2 R_{\rm vir, \, gr}$ from the center of the system. About 1 Mpc western P63661 a quite massive group is found, with 32 members. The properties of this group are also listed in Tab.\ref{tab:groups}. P63661 is found at 4.8 R$_{\rm vir, \, gr}$ from the center of this massive group.  This group is at the center of a filament, which extends both towards NW and towards SE for several Mpc { and includes 51 galaxies}. 

Finally, P8721 is part of a four-member system, embedded in the center of an extended filament { of 35 galaxies}. It is at 0.3 R$_{\rm vir, \, gr}$ from the group center and the closest of the two other galaxies of the group is at 165 kpc. 

{ Just for reference, we note that 
in the control sample used here (Sec. \ref{sec:morph}) 13 out of the 14 field galaxies are either binary or single systems, supporting the scenario that the environment might indeed play a role. } { Based on the definition of filaments by \cite{Tempel2014_f}, 7/14 galaxies are in small filaments (less than 25 objects), while the others are at the boundaries of filamentary structures. }

To understand whether the perturbed morphology of the galaxies are the result of tidal interactions with their closest neighbors, we follow a crude approach that was already exploited by, e.g., \cite{Wolter2015, Merluzzi2016} and estimate the acceleration $a_{tid}$ produced by the closest neighbour on the ISM of the galaxy of interest and compare it with the acceleration from the potential of the galaxy itself, $a_{gal}$. Following \cite{Vollmer2005}, 
$$
\frac{a_{tid}}{a_{gal}} = \frac{M_{neighbour}}{M_{gal}} \left(\frac{r}{R}-1\right)^{-2}
$$
where R is the distance from the centre of the galaxy, r is the distance between the galaxies \citep{Vollmer2005}, and $\frac{M_{neighbour}}{M_{gal}}$ the stellar mass ratio. This formulation would require the true distance, that we obviously do not have, therefore we can only use the projected distance. In all the three cases, $\frac{a_{tid}}{a_{gal}}<<1$ and we can exclude tidal interactions with the other group members.

Another source of perturbation to the galaxy morphology might be their position within the filament. 
Galaxies in filaments are indeed expected to have a very different  experience from those in largely empty regions \citep{Bahe2013}. 

All the three galaxies analysed have the major axis more or less aligned to the filament they are embedded in. They could therefore either be flowing along the filament or crossing it perpendicularly. In filaments the IGM density is enhanced, rising  the possible ram pressure intensity. In particular, galaxies with shallow potential wells can provide a relatively small restoring force, and a  significant gas stripping can take place at typical gas densities and velocities \citep[e.g.,][]{Benitez2013}. The clouds we observe could therefore actually be the result of a galaxy crossing a filament. 

Analytically quantifying the effect of the filament is not straightforward and also simulations have never been able to quantify the impact of this environment on the spatially resolved properties of the galaxies. 
Accurately measure the density of the IGM in these environments and estimate the 3D center of the filament needed to quantify the distance of the galaxy from it is indeed a quite hard task. It is however tantalizing to suppose filaments are responsible for the observed gas distribution.
We might therefore be witnessing the cosmic web stripping acting on galaxies more massive than dwarfs \citep{Benitez2013}.

However, stripping requires a relatively high velocity difference between the galaxy and the filament and the galaxies simulated by \cite{Benitez2013} were low mass objects, while the galaxies discussed here have $\log(M_\ast/M_\odot)\gtrsim 10$. So,  rather than stripping, we are most likely seeing gas compression due to the flowing of the galaxies within the filaments. This compression can be induced by an increase in surrounding thermal pressure and can switch on the surrounding clouds. Numerical simulations by \cite{Liao2018} show that filaments can assist  the gas cooling and increase the star formation in their residing dark matter haloes.
As a consequence, it might be possible that the densest regions in the circumgalactic gas get switched on in their star formation when the galaxy impacts with the sparse IGM. The detached clouds observed around the galaxies with no preferential orientation might be indeed an evidence for this phenomenon, that we call ``Cosmic web enhancement''.  In  \citet[][Paper XII]{Vulcani2018b} we have  presented another galaxy with similar features and found in a similar environment. 

Nonetheless, there are no simulations specifically focusing on the spatial properties of galaxies in filaments. Developing this kind of simulations is now urgent to better investigate this peculiar environment and its effect on galaxies.

Indeed,  different conditions of the filaments (density, extent, orientation), as well as the inclination of the galaxy with respect to the filament itself, could also have different impacts on the embedded galaxies, and this could explain the differences observed in P8721 with respect to the other two galaxies. 

\subsection{External  H{\sc ii} regions in the literature}\label{sec:liter}

{ As mentioned in the Introduction, in the literature few studies have identified \Ha knots at large radii \citep{Kennicutt1989, Martin2001, Ferguson1998} or isolated \citep{Gerhard2002, Cortese2004, RyanWeber2004, Oosterloo2004, Sakai2002, Mendes2004}. } 

All of above studies are based on traditional observational techniques  to observe the \Ha emission in the outskirts of galaxies. The most exploited one is narrow-band imaging with subsequent subtraction of broad-band continuum emission. This is however generally insufficiently sensitive to probe large radii. The limitation lies both in the achievable signal-to-noise ratio (S/N) and  in the stellar continuum subtraction. Higher spectral resolution is generally preferable  and very narrow filter bandpasses have also been adopted, as also traditional spectroscopy, that however can have quite low throughput. Since the pioneering work of Bland-Hawthorn et al. (1997), also the Fabry-Perot staring technique 
has been used. 

These techniques however only permit the detection and basic characterization of the H{\sc ii} regions, without giving spatially resolved information on the chemical composition and age of the regions. This is now possible thanks to the recent advent of Integral Field Spectrographs (IFS). However, the known on-going large IFS surveys like the Calar Alto Legacy Integral Field Area (CALIFA) Survey \citep{Sanchez2012}, the Sydney-AAO Multi-object Integral field spectrograph (SAMI) Survey \citep{Croom2012}, the Mapping Nearby Galaxies at Apache Point Observatory  (MaNGA) Survey \citep{Bundy2015} typically reach out to 2.5-3 effective radii  at most \citep{Bundy2015}, therefore are not designed to catch dis-homogeneities in the ionised gas distribution in the galaxy outskirts. 

GASP has been instead designed to focus on the galaxy external regions and allows us to perform a detailed analysis of the galaxy outskirts.

{ To put our results in context, }
we have directly compared the observed \Ha distributions of P95080, P19482, P63661 and P8721 to those of many other literature results, and confirmed that they present many peculiarities, hardly found in previous studies. 
Galaxies characterised by narrow band image surveys like  the \Ha Galaxy Survey (\Ha GS, Shane et al. 2002), the \Ha galaxy survey \citep{James2004}, the H-alpha Galaxy Groups Imaging Survey (H$\alpha$ggis, PI. Erwin), An \Ha Imaging Survey of Galaxies in the Local 11 Mpc Volume \cite[11Hugs][]{Kennicutt2008}, Dynamo \citep{Green2014}, or  Fabry-Perot observations like the Gassendi \Ha survey of SPirals \cite[GHASP][]{Epinat2008} almost never present such extended and luminous ($\rm{\log (H\alpha [erg/s/cm^2/arcsec^2]>-17.5}$) \Ha regions located well beyond R25. This might be due to the shallower surface brightness reached:  the surface brightness limit of the observations of \cite{James2004} is SB(\Ha${\rm +[N]) = 10^{-15}erg/cm^2/s/arcsec^2}$. In addition, at least some of these surveys \citep[e.g.,][]{Epinat2008} targeted galaxies in the cluster environment, such as Virgo, and those observed features are most likely due to the ram pressure exerted by the hot intracluster medium \citep{Gunn1972}.

The detached \Ha regions we detect are 
quite bright and large and we can exclude the galaxies are found in clusters. 

These regions are also  much brighter than the emission produced by the gaseous haloes and are similar to the intergalactic H{\sc ii} regions discovered by \cite[e.g.][]{RyanWeber2004} in terms of \Ha luminosity. However, the latter are not always bound to the main galaxy, while all the clouds we discussed present compatible velocities and ionised gas and stellar properties. \cite{RyanWeber2004} results are consistent with stars forming in interactive debris as a result of cloud-cloud collisions, while no signs of interactions are evident from our analysis. 

Unfortunately, no high resolution UV data are currently available for the three galaxies. They have been observed with GALEX, but these data are too shallow to detect any detached material.

\section{Conclusions}

GASP (GAs Stripping phenomena in galaxies with MUSE) is an ESO Large Program with the MUSE/VLT to study the causes and the effects of gas removal processes in galaxies in different environments in the local universe. Within the sample, we identified { four} galaxies that show peculiar ionised gas distributions: several \Ha clouds have been observed { beyond 4 $r_e$}. 
The gas kinematics, metallicity map and the ratios of emission line fluxes (BPT diagrams) confirm that they do belong to the galaxy gas disk, the stellar kinematics shows that very weak stellar continuum is associated to them. Similarly, the star formation history and luminosity weighted age maps point to a recent formation of such clouds, as also of more than half of the stellar disk for P95080, { P19482} and P63661.  The clouds are powered by star formation, and are characterised by intermediate values of extinction ($A_V\sim 0.3-0.5$). These, along with an object discussed in \citetalias{Vulcani2018b}, are the only three galaxies in all the GASP non cluster sample showing such tattered \Ha distribution, and we have not found any similar object in the currently existing literature surveys. 

The three galaxies share a similar location in the Universe: they all  belong to filamentary structures, therefore we point to a scenario in which the observed features are due to ``Cosmic web enhancement'': we hypothesize that we are witnessing galaxies passing through or flowing within filaments that are able to cool the gas and increase the star formation in the densest regions in the circumgalactic gas.  Observed differences among the three galaxies might be due to the different conditions of the filaments.  
\cite{Liao2018} showed that filaments are an environment that particularly favors this gas cooling followed by condensation and star formation enhancement.  
 
In the recent years, there has been an increasing interest for the role of filaments in affecting galaxy properties, nonetheless, to our knowledge, this paper presents the first analysis of the effect of this environment on the spatially resolved properties of the galaxies, highlighting the importance of this kind of data to get insights on galaxy evolution as a function of environment. Targeted simulations illustrating the effect of filaments on galaxy properties are now crucial to make progress on the physical processes acting in the different environments. 

\section*{Acknowledgements}

Based on observations collected at the European Organisation for Astronomical Research in the Southern Hemisphere under ESO programme 196.B-0578. We acknowledge funding from the INAF PRIN-SKA 2017 program 1.05.01.88.04 (PI Hunt). We acknowledge financial contribution from the contract ASI-INAF n.2017-14-H.0  Y.~J. acknowledges support from CONICYT PAI (Concurso Nacional de Inserci\'on en la Academia 2017) No. 79170132  and FONDECYT Iniciaci\'{o}n 2018 No. 11180558.



\bibliographystyle{mnras}
\bibliography{gasp} 







\bsp	
\label{lastpage}
\end{document}